\documentclass[twocolumn,prd,superscriptaddress,showpacs]{revtex4}
\usepackage{graphicx}
\usepackage{dcolumn}
\usepackage{amsmath}
\usepackage{simplewick}

\begin{document}

\title{Scattering of low energy neutrinos and antineutrinos by atomic electrons}
\author{Ian B. Whittingham}
\affiliation{College of Science and Engineering,
James Cook University, Townsville, Queensland, Australia 4811}

\date{\today}

\begin{abstract}
Studies of neutrino mixing and oscillations, solar neutrinos as background
in dark matter searches involving electron detection, detection of sterile neutrino 
warm dark matter, and of possible electromagnetic properties of neutrinos, 
have generated interest in the low energy O(10 keV)
scattering of electron neutrinos and antineutrinos by atomic electrons where the binding of the 
atomic electron cannot be ignored. Of particular interest is the ionization of atoms by neutrinos 
and antineutrinos. Most existing calculations are based upon modifications of the free electron
differential cross section which destroy the relationship between the neutrino helicities and 
the orbital and spin angular momenta of the atomic electrons. 
The present calculations maintain the full collision 
dynamics by formulating the scattering  in configuration space using the Bound Interaction 
Picture, rather than the usual formulation in the Interaction Picture in momentum space as 
appropriate to scattering by free electrons.
Energy spectra of ionization electrons produced by scattering of neutrinos and antineutrinos
with energies of 5, 10, 20, and 30 keV by hydrogen, helium and neon have been calculated 
using Dirac central field eigenfunctions, and are presented as ratios to
the spectra for scattering by free electrons.
Binding effects increase strongly with atomic number, are largest for low neutrino energy and, 
for each neutrino energy, greatest at the high electron energy end of the spectrum. The most
extreme effects of binding are for 5 keV scattering by Ne where the ratios are less
than $0.1$. The energy spectra have been calculated for both a Coulombic final electron state 
and a free final electron state.
The results indicate that the binding effects from the continuum state of the final electron are 
significant and can be comparable to those arising from the bound initial electron state. 
\end{abstract}

%\pacs{32.70.Jz, 34.50.Cx, 34.50.Rk, 34.20.Cf}
\maketitle

\section{Introduction}

Studies of neutrino mixing and oscillations~\cite{Vergados2010}, solar neutrinos as background
in dark matter searches involving electron detection~\cite{Thomas2016}, detection 
of sterile neutrino warm dark matter~\cite{Campos2016}, and of possible electromagnetic
properties of neutrinos, such as magnetic and electric dipole moments, using low energy elastic
scattering of neutrinos and antineutrinos~\cite{Giunti2015,Jeong2021}, have generated interest in the low energy O(10 keV)
scattering of electron neutrinos and antineutrinos by atomic electrons
\begin{equation}
\label{W0a}
\nu_{e} (\bar{\nu}_{e}) + e^{-} \rightarrow \nu_{e} (\bar{\nu}_{e}) + e^{-} .
\end{equation}
The standard scattering is due to the weak interaction and involves both $W$-boson 
charged current and $Z$-boson neutral current exchange. 
If neutrinos do have electromagnetic properties, generated by quantum loop effects, 
there will also be scattering due to single photon exchange. The weak and electromagnetic 
scatterings are incoherent and their dependences upon the energy transferred 
$T=E_{\nu_{i}}-E_{\nu_{f}}$ to the atomic electron  for $T \ll E_{\nu_{i}}$ are quite 
different, with their differential cross sections $d \sigma /dT$ being approximately 
constant for the standard scattering and $\propto 1/T$ for the electromagnetic 
scattering. The effect of a neutrino magnetic moment is then a distortion in the shape 
of the atomic electron recoil spectrum at low $E_{\nu_{i}}$.
For all these low energy studies the binding of the atomic electron cannot be ignored
and one can expect modifications of the free electron scattering formulae.
Neutrino-atom collisions has been reviewed by Kouzakov and Studenikin~\cite{Kouzakov2014}.
In this present study we consider only the scattering by the standard weak interaction.

Of particular interest is the ionization of atoms by neutrinos and antineutrinos. The case
of ionization of hydrogen-like atoms was first considered by~\cite{Gaponov1976} who found the 
ionization cross section per electron exceeded the free electron cross section by a factor
of 2 or 3 for neutrino energies $E_{\nu} \sim \alpha Z m_{e}$. Subsequently, calculated
electron spectra from inelastic scattering of neutrinos by atomic electrons of ${}^{19}$F and
${}^{96}$Mo were found~\cite{Fayans1992,Dobretsov1992} to differ significantly from 
scattering by a free electron and were always smaller than the free electron case.
Ionization cross sections for scattering by bound electrons in the light atoms H, He and Ne
were also found~\cite{Gounaris2002} to be smaller than the corresponding free electron
cross sections. The calculations were then extended~\cite{Gounaris2004} to the electron
spectra for H, He and Ne, and integrated ionization cross sections for H, He, Ne and Xe.

The calculations of~\cite{Gounaris2002,Gounaris2004} are based upon the assumption of spin-independent
non-relativistic atomic wave functions and consider the scattering to occur from a free electron
whose energy $E_{e_{i}}$ is set to the energy of the initial bound electron $m_{e}+\epsilon$, 
where $\epsilon$ is the binding energy, and whose momentum $\mathbf{p}_{e_{i}}$ is determined
by the probability amplitude $|\Psi_{n_{i}l_{i}m_{i}}(\mathbf{p}_{e_{i}})|^{2}$, where
$\Psi_{n_{i}l_{i}m_{i}}(\mathbf{p}_{e_{i}})$ is the momentum-space atomic wave function. 
The bound electron is then described by the effective squared mass
\begin{equation}
\label{W03a}
\tilde{m}^{2} = p_{e_{i}}^{2} = E_{e_{i}}^{2}-\mathbf{p}_{e_{i}}^{2}.
\end{equation}
Coulombic effects on the final electron are also ignored.
This allows the $\nu_{e}$-electron scattering process to be described as a probability weighted 
scattering by a free electron of mass $\tilde{m}$. This scattering can then be 
averaged (summed) over all initial (final)
electron and neutrino spin states, giving the invariant squared scattering amplitude
\begin{eqnarray}
\label{W01}
|F(\nu_{e}e^{-} \rightarrow \nu_{e}e^{-})|^{2} & = & 2 G_{F}^{2}
\{(\bar{v}_{e}-\bar{a}_{e})^{2} (s-m_{e}^{2}) (s-\tilde{m}^{2})
\nonumber  \\
&& + (\bar{v}_{e}+\bar{a}_{e})^{2}(u-m_{e}^{2})(u-\tilde{m}^{2})
\nonumber  \\
&& +2 m_{e}^{2} (\bar{v}_{e}^{2}-\bar{a}_{e}^{2}) t \},
\end{eqnarray}
where 
\begin{equation}
\label{W02}
s =(p_{\nu_{i}}+p_{e_{i}})^{2}, \quad
t=(p_{\nu_{i}}-p_{\nu_{f}})^{2}, \quad
u=(p_{\nu_{i}}-p_{e_{f}})^{2},
\end{equation}
are the usual kinematic invariants, and it has been assumed that the scattering 
occurs at low momentum transfers $t^{2} \ll M_{Z,W}^{2}$.
Here, $\bar{v}_{e}=1+4 \sin ^{2} \theta_{W}$ and $\bar{a}_{e} =-1$ where $\theta_{W}$ is the 
weak mixing angle.
For scattering by a free electron, $\tilde{m}^{2}$ is replaced by $m_{e}^{2}$.
The result for $\bar{\nu_{e}}$ - electron scattering follows from (\ref{W01}) by interchanging 
$s$ and $u$. The differential cross section in this approach is
\begin{eqnarray}
\label{W03}
d \sigma & = & \frac{1}{16 \pi^{2} E_{\nu_{i}}E_{e_{i}}} \delta^{(4)} (p_{e_{f}}+p_{\nu_{f}}-p_{e_{i}}-p_{\nu_{i}})
\nonumber  \\
&& \times |\Psi_{n_{i}l_{i}m_{i}}(\mathbf{p}_{e_{i}})|^{2}\frac{d^{3}\mathbf{p}_{e_{i}}}{(2 \pi )^{3}}
\nonumber  \\ 
&& \times |F(\nu_{e}e^{-} \rightarrow \nu_{e}e^{-})|^{2}
\frac{d^{3}\mathbf{p}_{\nu_{f}}}{2E_{\nu_{f}}} \frac{d^{3}\mathbf{p}_{e_{f}}}{2E_{e_{f}}}.
\end{eqnarray}

An alternative approach has been introduced ~\cite{Voloshin2010,Kouzakov2011a,Kouzakov2011b}
which assumes $T \ll E_{\nu_{i}}$ and $T \ll m_{e}$ so that the electrons and scattering can 
be treated non-relativistically. The atomic target is considered to be unpolarized. The
differential cross section is then the low-$T$ form of the free electron result modified to
the form
\begin{equation}
\label{W04}
\frac{d\sigma}{dT} = \frac{G_{F}^{2}}{4 \pi } \left(1 + 4 \sin^{2} \theta_{W} 
+ 8 \sin ^{4} \theta _{W} \right) \int  S(T,q^{2}) dq^{2} ,
\end{equation}
where $S(T,q^{2})$ is the dynamical structure function 
\begin{equation}
\label{W05}
S(T,q^{2}) = \sum_{f} \delta(T-E_{f}+E_{i}) 
|\langle f|\rho(\mathbf{q})|i \rangle |^{2},
\end{equation}
and $\mathbf{q}$ is the spatial momentum transfer with $T^{2} \le q^{2} \le 4 E_{\nu_{i}}^{2}$. 
The sum is over all final atomic states 
$|f \rangle $  of energy $E_{f}$ consistent with energy conservation, 
with $|i \rangle$ being the initial state. Here 
\begin{equation}
\label{W06}
\rho (\mathbf{q}) = \sum_{a=1}^{Z} \exp (i \mathbf{q} \cdot \mathbf{r}_{a})
\end{equation}
is the Fourier transform of the electron number density and the sum is over the 
positions $\mathbf{r}_{a}$  of all the $Z$ electrons in the atom.

The dynamical structure function is evaluated through its relationship 
\begin{equation}
\label{W07}
S(T,q^{2}) = \frac{1}{\pi} \mathrm{Im} F(T,q^{2}).
\end{equation}
to the density-density Green's function
\begin{equation}
\label{W08}
F(T,q^{2})= \sum_{f} \frac{|\langle f|\rho(\mathbf{q})|i \rangle |^{2}}{T-E_{f}+E_{i}-i \epsilon}.
\end{equation}
Atomic binding deforms the density-density Green's function by
broadening and shifting the free electron $\delta$-peak at $q^{2}=2m_{e}T$, but
Kouzakov \textit{et al}~\cite{Kouzakov2011b} argue that the modifications relative to 
the free -electron expressions are quite small. Analytical results are obtained for 
$1s$, $2s$ and $2p$ hydrogenlike states.

The more recent calculations by Chen \textit{et al.}~\cite{Chen2013,Chen2014,Chen2015} use the
four-fermion contact form for the weak interaction
\begin{equation}
\label{W09a}
\frac{d^{2} \sigma}{dT d\Omega} = \frac{G_{F}^{2}}{2 \pi^{2}} \frac{E_{\nu_{f}}}{E_{\nu_{i}}}  
L^{(\nu)\alpha \beta} R^{(w)}_{\alpha \beta}
\end{equation}
where the scattering of the neutrino of momentum $p_{\nu_{i}}$ and helicity $s_{i}$ 
is described by the tensor
\begin{eqnarray}
\label{W09b}
L^{(\nu)\alpha \beta} & = & 
\langle p_{\nu_{f}},s_{f}|\gamma^{\alpha}(1-\gamma_{5}) |p_{\nu_{i}},s_{i} \rangle  
\nonumber  \\
&& \times 
\langle p_{\nu_{f}},s_{f}|\gamma^{\beta}(1-\gamma_{5}) |p_{\nu_{i}},s_{i} \rangle ^{*}.
\end{eqnarray}
The effects on the atomic system are represented by the response functions
\begin{eqnarray}
\label{W09}
R^{(w)}_{\alpha \beta } & = & \frac{1}{2j_{i}+1}\sum_{m_{j_{i}}} \sum_{f} 
\langle f |j^{\alpha}_{w}|i \rangle \langle f | j^{\beta}_{w} |i \rangle ^{*} \nonumber \\
&& \times \delta (T+E_{i}-E_{f}),
\end{eqnarray}
which involve a sum/integral over the final atomic electron states $|f \rangle$ and a spin average over
the initial atomic states $|i \rangle = |j_{i},m_{j_{i}}, \ldots \rangle$. The relativistic
weak current representing the sum of the charged and neutral currents is
\begin{equation}
\label{W10a}
j^{\alpha}_{w}=\frac{1}{2}\bar{e}^{\prime}(\bar{v}_{e} \gamma^{\alpha} + 
\bar{a}_{e} \gamma^{\alpha}\gamma_{5} ) e.
\end{equation}
The model has been applied to Ge with the response functions evaluated using the multiconfiguration
relativistic random phase approximation. Consequently, the leading relativistic terms in the atomic 
Hamiltonian are treated nonperturbatively by using Dirac eigenfunctions, the two possible 
configurations for the Ge ground state are included, and the Random Phase Approximation 
accounts for the two-body correlations. The weak current operator (\ref{W10a}) is 
expanded in spherical multipoles.

The calculations~\cite{Gounaris2002,Gounaris2004,Voloshin2010,Kouzakov2011a,Kouzakov2011b}
destroy the relationship between the neutrino helicities and the orbital and spin 
angular momenta of the atomic electrons. Some of these issues are addressed by~\cite{Chen2014,Chen2015},
and their approach is closest in spirit to the present calculations.
To maintain the full collision dynamics, the scattering
of the neutrino by the bound electron will be treated in a similar manner to that of bound 
Compton scattering~\cite{Whittingham1971} in that the scattering will be formulated in 
configuration space using the Furry Bound Interaction Picture~\cite{Furry1951} rather than the usual formulation
in the Interaction Picture in momentum space as appropriate to scattering by free electrons. 

The general formalism for the scattering of neutrinos and antineutrinos by atomic electrons is 
presented in Sec. II. This includes the derivation of the $S$-matrix and differential cross sections 
for the scattering processes in terms of the contraction of neutrino and atomic electron tensor amplitudes, 
and the explicit evaluation of the atomic electron amplitude for the
case of an atomic electron represented by a central field Dirac eigenfunction. The nature of the 
radial matrix elements which occur in the atomic electron amplitude are discussed in Sec. III,
and issues relating to the evaluation of the cross sections in Sec. IV. Results for the energy spectra 
of  the ionization electrons produced by scattering of neutrinos and antineutrinos off hydrogen,
helium and neon are presented and discussed in Sec.V. Section VI contains a summary and conclusions
for the investigation. Details of the derivation of the $S$-matrix in the Bound Interaction Picture 
are given in Appendix A, explicit expressions for the electron scattering tensors in Appendix B, and
computational details for the evaluation of the radial matrix elements in Appendix C.

\section{General formalism}
\subsection{$S$-matrix for scattering by bound electrons}

As discussed above, in order to treat the effects of atomic binding on the scattering of neutrinos 
by atomic electrons, the second-order $S$-matrix element will be developed in the Furry 
Picture~\cite{Furry1951} in which the
electron is in the presence of a c-number electromagnetic field $A^{(\mathrm{ext})}_{\alpha}(x)$ and
the electron field operator satisfies
\begin{equation}
\label{W0}
[i \gamma^{\alpha}\partial_{\alpha} -e \gamma^{\alpha}A^{(\mathrm{ext})}_{\alpha}(x)-m_{e}]e(x) =0,
\end{equation}
where $\partial_{\alpha}\equiv \partial/\partial x^{\alpha}$.
The natural unit system $\hbar = c=1$ is used throughout, the 
scalar product of two 4-vectors is $A\cdot B \equiv g^{\alpha \beta} A_{\alpha} B_{\beta} 
= A_{0}B_{0}-\mathbf{A}\cdot \mathbf{B}$, the Dirac matrices 
$\gamma^{\alpha}, (\alpha =0,1,2,3)$ satisfy $\{\gamma^{\alpha},\gamma^{\beta}\}=2g^{\alpha \beta}$,
$\gamma_{5}\equiv i\gamma^{0}\gamma^{1}\gamma^{2}\gamma^{3}$, 
and the field operators for a given particle are denoted by the symbol for that particle.

The part of the Standard Model lepton interaction Lagrangian which describes the interactions between
an electron-neutrino $\nu_{e}$ and an electron $e$ is~\cite{IZ80,Bailin82}
\begin{equation}
\label{W1}
\mathcal{L}^{\nu_{e}e}_{I} =
\mathcal{L}^{\nu_{e}We}_{I}+\mathcal{L}^{\nu_{e}Z\nu_{e}}_{I}+\mathcal{L}^{eZe}_{I},
\end{equation}
where
\begin{eqnarray}
\label{W2}
\mathcal{L}^{\nu_{e}We}_{I}& = & \frac{-g}{2 \sqrt{2}} 
N[\bar{\nu_{e}} \gamma^{\alpha} (1-\gamma_{5})W^{(+)}_{\alpha} e \nonumber  \\
&& +\bar{e} \gamma^{\alpha} (1-\gamma_{5}) W^{(-)}_{\alpha} \nu_{e} ] ,  \\
\mathcal{L}^{\nu_{e}Z\nu_{e}}_{I} & = & \frac{-g}{4 \cos \theta_{W}} 
N[\bar{\nu_{e}} \gamma^{\alpha} (1-\gamma_{5})Z_{\alpha} \nu_{e} ], \\
\mathcal{L}^{eZe}_{I} & = & \frac{-g}{4 \cos \theta_{W}} 
N[\bar{e} \gamma^{\alpha} (v_{e}+a_{e}\gamma_{5}) Z_{\alpha} e ].
\end{eqnarray}
Here $g$ is the $SU(2)$ gauge coupling constant, $v_{e}=-1+4 \sin^{2} \theta_{W}$ 
and $a_{e}=1$ are the weak neutral current parameters,
$W^{(\pm)}$ and $Z$ are the charged and neutral weak gauge boson field operators respectively, 
and $N$ is the normal ordering operator. 

The total $S$-matrix for $\nu_{e}$ scattering at low momentum transfers $k^{2} \ll M_{A}^{2}$, 
where $A=W,Z$, is (see Appendix A)  
\begin{equation}
\label{W34}
S_{fi}^{(\nu)}= -\pi i\frac{G_{\mathrm{F}}}{\sqrt{2}} \delta(E_{fi}^{(\nu)}) 
M^{(e)}_{n_{f},n_{i}}(\mathbf{q})^{\alpha} M^{(\nu)}(\mathbf{p}_{\nu_{f}},s_{f},\mathbf{p}_{\nu_{i}},s_{i})_{\alpha}
\end{equation}
where
\begin{eqnarray}
\label{W35}
M^{(e)}_{n_{f},n_{i}}(\mathbf{q})^{\alpha} & = & \int d^{3}x \;
e^{i(\mathbf{p}_{\nu_{i}}-\mathbf{p}_{\nu_{f}})\cdot \mathbf{x}}  
\nonumber  \\
&& \times
\bar{\phi}^{(+)}_{n_{f}} (\mathbf{x}) \gamma^{\alpha}(\bar{v}_{e}+\bar{a}_{e}\gamma_{5})
\phi^{(+)}_{n_{i}}(\mathbf{x}),
\end{eqnarray}
and
\begin{equation}
\label{W36}
M^{(\nu)}(\mathbf{p}_{\nu_{f}},s_{f},\mathbf{p}_{\nu_{i}},s_{i})_{\alpha} = \bar{u}^{(s_{f})}(\mathbf{p}_{\nu_{f}})\gamma_{\alpha}
(1-\gamma_{5}) u^{(s_{i})}(\mathbf{p}_{\nu_{i}}).
\end{equation}
Here $u^{(s)}(\mathbf{p}_{\nu})$ are the plane wave spinors describing a neutrino with momentum $\mathbf{p}_{\nu}$
and helicity $s$, and $\phi^{(+)}_{n}(\mathbf{x})$ is the energy eigenfunction for an electron in a 
state of the external field $A^{(\mathrm{ext})}$ specified by the quantum numbers $n$. The quantity
\begin{equation}
\label{W29a}
\delta(E_{fi}^{(\nu)}) \equiv \delta(E_{n_{f}}+E_{\nu_{f}}-E_{n_{i}}-E_{\nu_{i}}) .
\end{equation}
incorporates energy conservation, and $\mathbf{q}=\mathbf{p}_{\nu_{i}}-\mathbf{p}_{\nu_{f}}$ is 
the momentum transfer from the neutrino.
The electron mixing parameters are
\begin{equation}
\label{W37}
\bar{v}_{e} = v_{e}+2 = 1+4 \sin^{2} \theta_{W}, \quad \bar{a}_{e}= a_{e}-2 = -1.
\end{equation}
For scattering of antineutrinos, $M^{(\nu)}$ is replaced by
\begin{equation}
\label{W36a}
M^{(\bar{\nu})}(\mathbf{p}_{\nu_{f}},s_{f},\mathbf{p}_{\nu_{i}},s_{i})_{\alpha} = 
\bar{v}^{(s_{i})}(\mathbf{p}_{\nu_{i}})\gamma_{\alpha}
(1-\gamma_{5}) v^{(s_{f})}(\mathbf{p}_{\nu_{f}}),
\end{equation}
where $v^{(s)}(\mathbf{p}_{\nu})$ is the antineutrino plane wave spinor,
and $\delta(E_{fi}^{(\nu)})$ is replaced by $\delta(E_{fi}^{(\bar{\nu})})$.

\subsection{Cross section}

We assume each atomic electron acts as an independent scattering center.
In order to obtain the scattering cross section per atomic electron, $S_{fi}^{(\nu)}$ is expressed in the form
\begin{equation}
\label{W38}
S_{fi}^{(\nu)} = \delta(E_{n_{f}}+E_{\nu_{f}}-E_{n_{i}}-E_{\nu_{i}}) \mathcal{M}_{fi}^{(\nu)}.
\end{equation}
The corresponding transition probability per unit time is then~\cite{Jauch55}
\begin{equation}
\label{W39}
dP_{fi}^{(\nu)} = \frac{1}{2 \pi } \delta(E_{n_{f}}+E_{\nu_{f}}-E_{n_{i}}-E_{\nu_{i}})|\mathcal{M}_{fi}^{(\nu)}|^{2}.
\end{equation}
For $\nu_{e}$ scattering into the momentum interval 
$(\mathbf{p}_{\nu_{f}}, \mathbf{p}_{\nu_{f}}  + d^{3}\mathbf{p}_{\nu_{f}})$, the transition probability per unit time is
\begin{equation}
\label{W40}
dP^{(\nu)} = \sum_{n_{f}} dP_{fi}^{(\nu)} \, d\rho_{f}^{(\nu)} d\rho_{f}^{(e)},
\end{equation}
where $d \rho_{f}^{(\nu)}$ ( $d \rho_{f}^{(e)}$)  is the density of final $\nu_{e}$ ($e$) states 
and the sum is over all final electron states consistent with energy conservation.  
For plane wave neutrino spinors 
normalized to $u^{\dagger(s)}(\mathbf{p}_{\nu}) u^{(s)}(\mathbf{p}_{\nu})=2 E_{\nu}$, the density of 
states is $d\rho_{f}^{(\nu)}=d^{3}\mathbf{p}_{\nu_{f}}/[(2 \pi)^{3}2E_{\nu_{f}}]$ and 
the incident neutrino flux is $2 E_{\nu_{i}}$.
 
The differential cross section is then
\begin{equation}
\label{W41}
d \sigma^{(\nu)} = \frac{dP^{(\nu)}}{2 E_{\nu_{i}}} = \frac{1}{2 E_{\nu_{i}}} \sum_{n_{f}} dP_{fi}^{(\nu)} \,
\frac{d^{3}\mathbf{p}_{\nu_{f}}}{ (2 \pi)^{3} 2E_{\nu_{f}}}\;d\rho_{f}^{(e)}.
\end{equation}
Writing $d^{3}\mathbf{p}_{\nu_{f}}=E_{\nu_{f}}^{2} d E_{\nu_{f}} d \Omega_{\nu_{f}}$ then
\begin{eqnarray}
\label{W42}
d \sigma^{(\nu)} & = & \frac{1}{(2 \pi)^{4} } \frac{1}{4 E_{\nu_{i}}}
\sum_{n_{f}}  \delta(E_{fi}^{(\nu)})
\nonumber  \\
&& \times  E_{\nu_{f}} d E_{\nu_{f}} d \Omega_{\nu_{f}} d\rho_{f}^{(e)}
 |\mathcal{M}_{fi}^{(\nu)}|^{2},  
\end{eqnarray}
where, from (\ref{W34}),
\begin{equation}
\label{W43}
\mathcal{M}_{fi}^{(\nu)} = - \pi i \frac{G_{\mathrm{F}}}{\sqrt{2}} 
M^{(e)}_{n_{f},n_{i}}(\mathbf{q})^{\alpha} M^{(\nu)}(\mathbf{p}_{\nu_{f}},s_{f},\mathbf{p}_{\nu_{i}},s_{i})_{\alpha}.
\end{equation}

The neutrino contribution to $|\mathcal{M}_{fi}^{(\nu)}|^{2}$ is~\cite{Bailin82}
\begin{eqnarray}
\label{W44}
L^{(\nu)}(p_{\nu_{i}},p_{\nu_{f}})^{\beta \alpha} &\equiv & 
[\bar{u}^{(s_{f})}(\mathbf{p}_{\nu_{f}})\gamma^{\beta}(1-\gamma_{5}) u^{(s_{i})}(\mathbf{p}_{\nu_{i}})]^{\dagger}
\nonumber  \\
&& \times \bar{u}^{(s_{f})}(\mathbf{p}_{\nu_{f}})\gamma^{\alpha}(1-\gamma_{5}) u^{(s_{i})}(\mathbf{p}_{\nu_{i}})
\nonumber \\
& = & 8 (p_{\nu_{i}}^{\beta}\,p_{\nu_{f}}^{\alpha} + p_{\nu_{i}}^{\alpha}\, p_{\nu_{f}}^{\beta} 
- p_{\nu_{i}} \cdot p_{\nu_{f}}\, g^{\beta \alpha}
\nonumber  \\
&& + i \epsilon^{\rho \beta \lambda \alpha}\, p_{\nu_{i,\rho}}\, p_{\nu_{f, \lambda}})  \label{W44a},
\end{eqnarray}
where $s_{i}=s_{f}=-1/2$.
The scattering of antineutrinos involves
\begin{eqnarray}
\label{W44a}
L^{(\bar{\nu})}(p_{\nu_{i}},p_{\nu_{f}})^{\beta \alpha} &\equiv & 
[\bar{v}^{(s_{i})}(\mathbf{p}_{\nu_{i}})\gamma^{\beta}(1-\gamma_{5}) v^{(s_{f})}(\mathbf{p}_{\nu_{f}})]^{\dagger}
\nonumber  \\
&& \times \bar{v}^{(s_{i})}(\mathbf{p}_{\nu_{i}})\gamma^{\alpha}(1-\gamma_{5}) v^{(s_{f})}(\mathbf{p}_{\nu_{f}})
\nonumber \\
& = & L^{(\nu)}(-p_{\nu_{f}},-p_{\nu_{i}})^{\beta \alpha},  
\nonumber \\
& = & [L^{(\nu)}(p_{\nu_{i}},p_{\nu_{f}})^{\beta \alpha}]^{*},
\end{eqnarray} 
where $s_{i}=s_{f}=+1/2$.

\subsection{Atomic electron amplitude}

The electron amplitude (\ref{W35}) requires the solutions $\phi^{(+)}_{n}(x)$ of (\ref{W0}). We assume the 
atomic electron moves in a spherically symmetric potential $V(r)=e A^{(\mathrm{ext})}(r)$ 
and use the Dirac representation for the $\gamma $ matrices
\begin{equation}
\label{W50}
\gamma^{0} =\left( \begin{array}{cc}
I & 0 \\
0  & -I 
\end{array} \right),
\gamma^{k} = \left( \begin{array}{cc}
0 & \sigma^{k} \\
- \sigma^{k}  & 0 
\end{array} \right), 
\gamma_{5} =\left(  \begin{array}{cc}
0  & I  \\
I  & 0
\end{array}
\right),
\end{equation}
where  $\sigma^{k}, k=1,2,3,$ are the Pauli 2x2 matrices, and $I$ is the unit 2x2 matrix.
The eigenfunctions have the form~\cite{Rose1961}
\begin{equation}
\label{W45}
\phi_{\kappa, \mu,E}(r, \theta, \varphi ) = \frac{1}{r} \left( 
\begin{array}{l}
g_{\kappa,E}(r) \chi^{\mu}_{\kappa}(\Omega) \\
i f_{\kappa, E} (r) \chi^{\mu}_{-\kappa}(\Omega)
\end{array}
\right).
\end{equation}
where $(r, \theta, \varphi )=(r, \Omega)$ are spherical polar coordinates, and $\chi^{\mu}_{\kappa}(\Omega)$
are the spinor spherical harmonics
\begin{equation}
\label{W46}
\chi^{\mu}_{\kappa}(\Omega) = \sum_{m_{s}} C(l_{\kappa},\frac{1}{2},j,\mu-m_{s},m_{s},\mu ) 
Y^{\mu-m_{s}}_{l_{\kappa}}(\Omega) \chi_{m_{s}}.
\end{equation}
Here $C(j_{1},j_{2},j_{3},m_{1},m_{2},m_{3})$ is a Clebsch-Gordon coefficient, and
$\chi_{m_{s}}$ are the two component Pauli spinors. The total angular momentum $j$ and orbital angular 
momentum $l_{\kappa}$ are obtained from the quantum number $\kappa$ by
\begin{equation}
\label{W47}
j=|\kappa|-\frac{1}{2}, l_{\kappa}=\left\{ \begin{array}{cl}
\kappa & \kappa > 0  \\
-\kappa-1& \kappa < 0
\end{array}, \right.
l_{-\kappa} = l_{\kappa} - \frac{\kappa}{|\kappa|},
\end{equation}
where $\kappa$ takes all non-zero integral values. Note that the subscript $e$ has 
been dropped from the electron energies.
The radial functions satisfy
\begin{equation}
\label{W48}
\left( \begin{array}{cc}
d/dr + \kappa /r & -(E+m_{e}-V(r))  \\
E-m_{e}-V(r) & d/dr-\kappa /r
\end{array}
\right)
\left( \begin{array}{c}
g_{\kappa,E}(r) \\
f_{\kappa,E}(r) 
\end{array}
\right)  = 0.
\end{equation}

As no observation is made upon the final continuum electron, all possible states of the electron 
must be summed over, with the result that the asymptotic form of the continuum eigenfunction is
not important~\cite{Olsen1955}. Any set of continuum functions may be used and the form (\ref{W45})
is the obvious choice.

In order to evaluate
\begin{eqnarray}
\label{W49}
N_{fi}^{\alpha}(\mathbf{q}) & \equiv & \int r^{2}\, dr \,d \Omega \; e^{i\mathbf{q}\cdot \mathbf{r}} 
\bar{\phi}_{\kappa_{f},\mu_{f},E_{f}} (r,\theta, \varphi) 
\nonumber  \\
&& \times \gamma^{\alpha}(\bar{v}_{e}+\bar{a}_{e}\gamma_{5})
\phi_{\kappa_{i},\mu_{i},E_{i}}(r,\theta, \varphi), 
\end{eqnarray}
we introduce the expansion
\begin{equation}
\label{W49a}
e^{i\mathbf{q}\cdot \mathbf{r}} = 4 \pi \sum_{l=0}^{\infty} \sum_{m=-l}^{+l} i^{l} 
j_{l}(|\mathbf{q}|r) Y^{m}_{l}(\hat{\mathbf{q}})^{*} \,Y^{m}_{l} (\Omega)
\end{equation}
and note that
\begin{equation}
\label{W51}
\Gamma^{(\alpha)}  \equiv  \gamma^{0} \gamma^{\alpha}(\bar{v}_{e}+\bar{a}_{e}\gamma_{5}) 
\end{equation}
have the form
\begin{eqnarray}
\label{W51a}
\Gamma^{(0)} & = & \left( \begin{array}{cc} 
\bar{v}_{e} & \bar{a}_{e} \\
\bar{a}_{e} & \bar{v}_{e} 
\end{array}
\right), \\
\Gamma^{(k)} & = & \left( \begin{array}{cc} 
\bar{a}_{e} \sigma^{k} & \bar{v}_{e}\sigma^{k} \\
\bar{v}_{e} \sigma^{k} & \bar{a}_{e} \sigma^{k}
\end{array}
\right).
\end{eqnarray}
Hence (\ref{W49}) becomes
\begin{eqnarray}
\label{W52}
N_{fi}^{(0,k)} (\mathbf{q}) & = &  4 \pi \sum_{l,m} i^{l} Y^{m}_{l}(\hat{\mathbf{q}})^{*}
\nonumber  \\
&& \times \{[\bar{v}_{e} (\bar{a}_{e}) [I^{gg}_{l}(q) \langle \chi^{\mu_{f}}_{\kappa_{f}}|
Y^{m}_{l}(I,\sigma^{k})| \chi^{\mu_{i}}_{\kappa_{i}} \rangle 
\nonumber  \\  
&& +I^{ff}_{l}(q) \langle \chi^{\mu_{f}}_{-\kappa_{f}}|Y^{m}_{l}
(I,\sigma^{k})| \chi^{\mu_{i}}_{-\kappa_{i}} \rangle ] ]  \nonumber  \\
&& +i \bar{a}_{e}(\bar{v}_{e}) 
[I^{gf}_{l}(q) \langle \chi^{\mu_{f}}_{\kappa_{f}}|
Y^{m}_{l}(I,\sigma^{k})| \chi^{\mu_{i}}_{-\kappa_{i}} \rangle 
\nonumber  \\ 
&& -I^{fg}_{l}(q) \langle \chi^{\mu_{f}}_{-\kappa_{f}}|Y^{m}_{l}
(I,\sigma^{k})| \chi^{\mu_{i}}_{\kappa_{i}} \rangle ] \}  
\end{eqnarray}
where, e.g. $  \bar{v}_{e} (\bar{a}_{e})$ for $\alpha =0(k)$ respectively, 
$q \equiv |\mathbf{q}|$ and the radial integrals are
\begin{eqnarray}
\label{W53}
I^{gg}_{l}(q) &\equiv & \int dr \,g^{*}_{\kappa_{f},E_{f}}(r) j_{l}(qr) g_{\kappa_{i},E_{i}}(r),
\nonumber \\
I^{gf}_{l}(q) &\equiv & \int dr \,g^{*}_{\kappa_{f},E_{f}}(r) j_{l}(qr) f_{\kappa_{i},E_{i}}(r),
\nonumber  \\
I^{fg}_{l}(q) &\equiv & \int dr \,f^{*}_{\kappa_{f},E_{f}}(r) j_{l}(qr) g_{\kappa_{i},E_{i}}(r),
\nonumber \\
I^{ff}_{l}(q) &\equiv & \int dr \,f^{*}_{\kappa_{f},E_{f}}(r) j_{l}(qr) f_{\kappa_{i},E_{i}}(r).
\end{eqnarray}
Here on we use the simplified notation $l_{\kappa_{i,f}}=l_{i,f}$,
 $l_{-\kappa_{i,f}}=l^{\prime}_{i,f}$ and $m_{s_{i,f}}=m_{i,f}$.

The matrix elements of $\sigma^{k}$ can be evaluated by transforming to a spherical basis
$\sigma^{\lambda}, \lambda =0,\pm 1,$ where
\begin{equation}
\label{W55}
\sigma^{\pm 1} \equiv \mp \frac{1}{\sqrt{2}}(\sigma^{1} \pm i \sigma^{2}), \sigma^{0}=\sigma^{3}
\end{equation}
and using the Wigner-Eckart theorem
\begin{equation}
\label{W56}
\langle \chi_{m_{f}}|\sigma^{\lambda} | \chi_{m_{i}} \rangle =
C(\frac{1}{2}, 1, \frac{1}{2},m_{i}, \lambda ,m_{f}) 
\langle \frac{1}{2}|| \sigma || \frac{1}{2} \rangle ,
\end{equation}
where the reduced matrix element is 
\begin{equation}
\label{W57}
\langle \frac{1}{2}|| \sigma || \frac{1}{2} \rangle = \sqrt{3}.
\end{equation}
Similarly,
\begin{eqnarray}
\label{W58}
\langle Y^{\mu_{f}-m_{f}}_{l_{f}} | Y^{m}_{l} | Y^{\mu_{i}-m_{i}}_{l_{i}} \rangle & = &
C(l_{i},l,l_{f},\mu_{i}-m_{i}, m, \mu_{f}-m_{f} )  \nonumber  \\
&& \times \langle l_{f}|| Y_{l} || l_{i} \rangle ,
\end{eqnarray}
where
\begin{equation}
\label{W59}
\langle l_{f}|| Y_{l} || l_{i} \rangle = \frac{[l_{i}l]}{\sqrt{4 \pi} [l_{f}]}
C(l_{i},l,l_{f},0,0,0).
\end{equation}
Here $[ab \ldots]=[(2a+1)(2b+1) \ldots]^{1/2}$.
The matrix elements between the spinor harmonics in the spherical basis can then be written 
\begin{widetext}
\begin{equation}
\label{W60}
\langle \chi^{\mu_{f}}_{\kappa_{f}}|Y^{m}_{l}(I,\sigma^{\lambda})| \chi^{\mu_{i}}_{\kappa_{i}} \rangle 
 =  \frac{[l]}{\sqrt{4 \pi }}  
J^{(I,\lambda)}_{l,m}(\kappa_{f},\kappa_{i})
\end{equation}
where
\begin{eqnarray}
\label{W61}
J^{(I)}_{l,m}(\kappa_{f},\kappa_{i}) & = &  \frac{[l_{i}]}{[l_{f}]} C(l_{i},l,l_{f},0,0,0) \sum_{m_{f},m_{i}} 
C(l_{f},\frac{1}{2},j_{f}, \mu_{f}-m_{f},m_{f},\mu_{f}) C(l_{i},\frac{1}{2},j_{i}, \mu_{i}-m_{i},m_{i},\mu_{i})
\nonumber  \\
&& \times C(l_{i},l,l_{f}, \mu_{i}-m_{i},m,\mu_{f}-m_{f}) \delta_{m_{f},m_{i}} \delta_{m,\mu_{f}-\mu_{i}}
\nonumber  \\
& = & \delta_{m,\mu_{f}-\mu_{i}} [l_{i}j_{i}] C(l_{i},l,l_{f},0,0,0)   W(l,l_{i},j_{f},\frac{1}{2},l_{f},j_{i}) 
C(l,j_{i},j_{f},\mu_{f}-\mu_{i},\mu_{i},\mu_{f}),
\end{eqnarray}
and~\cite{Whittingham1971}
\begin{eqnarray}
\label{W62}
J^{(\lambda)}_{l,m}(\kappa_{f},\kappa_{i}) & = & \sqrt{3} \frac{[l_{i}]}{[l_{f}]}  C(l_{i},l,l_{f},0,0,0)  \sum_{m_{f},m_{i}} 
C(l_{f},\frac{1}{2},j_{f}, \mu_{f}-m_{f},m_{f},\mu_{f}) C(l_{i},\frac{1}{2},j_{i}, \mu_{i}-m_{i},m_{i},\mu_{i})
\nonumber  \\
&& \times C(l_{i},l,l_{f}, \mu_{i}-m_{i},m,\mu_{f}-m_{f}) C(\frac{1}{2},1,\frac{1}{2},m_{i},\lambda ,m_{f} )
\nonumber  \\
& = & \delta_{m,\mu_{f}-\mu_{i}-\lambda}  \sqrt{6}[l_{i}j_{i}] C(l_{i},l,l_{f},0,0,0)
\sum_{f} [f] W(l,l_{i},j_{f},\frac{1}{2},l_{f},f) 
W(1,\frac{1}{2},f,l_{i},\frac{1}{2},j_{i}) \nonumber  \\
&& \times C(l,f,j_{f},\mu_{f}-\mu_{i}-\lambda, \mu_{i}+\lambda, \mu_{f})
C(j_{i},1,f,\mu_{i},\lambda ,\mu_{i}+\lambda),
\end{eqnarray}
\end{widetext}
where $W(a,b,c,d,e,f)$ is a Racah coefficient. The elements with 
$-\kappa_{i}$ or $-\kappa_{f}$ are obtained from the above through the 
replacements $l_{i}\rightarrow l_{i}^{\prime}$ and 
$l_{f} \rightarrow l_{f}^{\prime}$ respectively.

For scattering by all of the electrons in a specified atomic shell or subshell 
(labelled by $\kappa_{i}$), with no reference being 
made to the final state of the atomic electron, the cross section must be summed over all
possible initial and final electron states. We therefore need the quantities
\begin{equation}
\label{W63}
L^{(e)}_{fi}(\mathbf{q})^{\beta \alpha }= \sum_{\kappa_{f},\mu_{f},\mu_{i}}
 N_{fi}^{\beta}(\mathbf{q})^{*} N_{fi}^{\alpha}(\mathbf{q}).
\end{equation}
For the case of H, the $\mu_{i}$ summation over electrons in each shell or subshell is replaced
by its average over the K-shell.

The calculation of these coefficients is greatly simplified by choosing the coordinate system such that
the neutrino momentum transfer $\mathbf{q}$ is along the $Oz$ axis and $\mathbf{p}_{\nu_{i}}$ lies 
in the $x-z$ plane. Thus
\begin{equation}
\label{W65}
Y^{m}_{l}(\hat{\mathbf{q}}) = \frac{[l]}{ (4 \pi )^{1/2}}\;\delta_{m,0}.
\end{equation}
For $m=\mu_{f}-\mu_{i}$ this gives $\mu_{f}=\mu_{i}$ whereas, for $m=\mu_{f}-\mu_{i}-\lambda$, this 
gives $\lambda =\mu_{f}-\mu_{i}$. 

We need the combinations
\begin{widetext}
\begin{eqnarray}
A^{(I,I)}(l_{1},l_{2},l_{3},l_{4}) & = & \sum_{\mu_{f},\mu_{i}}
J^{(I)}_{\bar{l},\bar{m}}(\pm \kappa_{f},\pm \kappa_{i})^{*} J^{(I)}_{l,m}(\pm \kappa_{f},\pm \kappa_{i}), 
\label{W64a}  \\
A^{(I,\lambda )}(l_{1},l_{2},l_{3},l_{4}) & = & \sum_{\mu_{f},\mu_{i}}
J^{(I)}_{\bar{l},\bar{m}}(\pm \kappa_{f},\pm \kappa_{i})^{*} J^{(\lambda)}_{l,m}(\pm \kappa_{f},\pm \kappa_{i}) ,
\label{W64b}  \\
A^{(\lambda, I)}(l_{1},l_{2},l_{3},l_{4}) & = & \sum_{\mu_{f},\mu_{i}}
J^{(\lambda)}_{\bar{l},\bar{m}}(\pm \kappa_{f},\pm \kappa_{i})^{*} J^{(I)}_{l,m}(\pm \kappa_{f},\pm \kappa_{i}),
\label{W64c}  \\
A^{(\lambda^{\prime},\lambda )}(l_{1},l_{2},l_{3},l_{4}) & = & \sum_{\mu_{f},\mu_{i}}
J^{(\lambda^{\prime})}_{\bar{l},\bar{m}}(\pm \kappa_{f},\pm \kappa_{i})^{*} J^{(\lambda)}_{l,m}(\pm \kappa_{f},\pm\kappa_{i}) .
\label{W64d}  
\end{eqnarray}
Here $(l_{1},l_{2},l_{3},l_{4}) =(l_{f},l_{i}, l_{f},l_{i})$ for the case $(+\kappa_{f},+\kappa_{i})$,
with the replacements $l_{i} \rightarrow l_{i}^{\prime}$ for the case $-\kappa_{i}$ and $l_{f} \rightarrow l_{f}^{\prime}$ 
for the case $-\kappa_{f}$. With $m=\bar{m}=0$, then
$\lambda =0$ in (\ref{W64b}, \ref{W64c}) and $\lambda^{\prime}=\lambda$ in (\ref{W64d}).

After some standard Racah algebra manipulations (see, e.g. \cite{Rose1957}) we obtain 
\begin{eqnarray}
\label{W66}
A^{(I,I)}_{\bar{l}l}(l_{1},l_{2},l_{3},l_{4}) & = & \delta_{\bar{l},l} [j_{f} j_{i}]^{2} 
\frac{[l_{2}l_{4}]}{[l]^{2}} 
C(l_{2},l,l_{1},0,0,0) C(l_{4},l,l_{3},0,0,0) W(l,l_{2},j_{f},\frac{1}{2},l_{1},j_{i})
W(l,l_{4},j_{f},\frac{1}{2},l_{3},j_{i}),  \\
A^{(I,\lambda)}_{\bar{l}l}(l_{1},l_{2},l_{3},l_{4}) & = & \delta_{\lambda, 0} (-1)^{j_{i}-j_{f}+l}
\sqrt{6} [j_{f} j_{i}]^{2} 
\frac{[l_{2}l_{4}]}{[l]} C(l_{2},\bar{l},l_{1},0,0,0) C(l_{4},l,l_{3},0,0,0)  C(\bar{l},1,l,0,0,0)  \nonumber  \\
&& \times W(\bar{l},l_{2},j_{f},\frac{1}{2},l_{1},j_{i}) 
\sum_{f} [f]^{2}  W(l,l_{4},j_{f},\frac{1}{2},l_{3},f) W(1,\frac{1}{2},f,l_{4},\frac{1}{2},j_{i})
W(\bar{l}, j_{i},l,f,j_{f},1), \\
A^{(\lambda,I)}_{\bar{l}l}(l_{1},l_{2},l_{3},l_{4}) & = & A^{(I,\lambda)}_{l\bar{l}}(l_{3},l_{4},l_{1},l_{2}),
\\
A^{(\lambda^{\prime},\lambda)}_{\bar{l}l}(l_{1},l_{2},l_{3},l_{4}) & = & \delta_{\lambda^{\prime},\lambda}
(-1)^{j_{i}-j_{f}-l+\bar{l}+\lambda} \;6 [j_{f}j_{i}]^{2} [l_{2}l_{4}]
 C(l_{2},\bar{l},l_{1},0,0,0) C(l_{4},l,l_{3},0,0,0) \nonumber  \\
&& \times \sum_{g} C(\bar{l},l,g,0,0,0) C(1,1,g,-\lambda, \lambda, 0) \sum_{\bar{f},f} [\bar{f}f]^{2}
W(\bar{l},l_{2},j_{f},\frac{1}{2},l_{1},\bar{f}) W(l,l_{4},j_{f},\frac{1}{2},l_{3},f)  \nonumber  \\
&& \times W(1,\frac{1}{2},\bar{f},l_{2},\frac{1}{2},j_{i}) W(1,\frac{1}{2},f,l_{4},\frac{1}{2},j_{i}) 
W(\bar{l},\bar{f},l,f,j_{f},g)  W(1,\bar{f},1,f,j_{i},g).
\end{eqnarray}

In order to evaluate the quantities (\ref{W63}) we need the Cartesian components 
$\tilde{A}^{\beta,\alpha}(l_{1234})$, where
\begin{eqnarray}
\label{W67}
\tilde{A}^{0,0}_{\bar{l}l}(l_{1234}) &= & A^{(I,I)}_{\bar{l}l}(l_{1234}), \nonumber  \\
\tilde{A}^{1,1}_{\bar{l}l}(l_{1234}) &= & \tilde{A}^{2,2}_{\bar{l}l}(l_{1234}) =  
\frac{1}{2} [A^{(1,1)}_{\bar{l}l}(l_{1234}) + A^{(-1,-1)}_{\bar{l}l}(l_{1234})],
\nonumber  \\
\tilde{A}^{1,2}_{\bar{l}l}(l_{1234}) &= & -\tilde{A}^{2,1}_{\bar{l}l}(l_{1234}) = 
- \frac{i}{2} [A^{(1,1)}_{\bar{l}l}(l_{1234}) - A^{(-1,-1)}_{\bar{l}l}(l_{1234})] ,
\nonumber  \\
\tilde{A}^{0,3}_{\bar{l}l}(l_{1234}) &= & A^{(I,0)}_{\bar{l}l}(l_{1234}), \quad
\tilde{A}^{3,0}_{\bar{l}l}(l_{1234}) =  A^{(0,I)}_{\bar{l}l}(l_{1234}),
\nonumber  \\
\tilde{A}^{3,3}_{\bar{l}l}(l_{1234}) &= & A^{(0,0)}_{\bar{l}l}(l_{1234}).
\end{eqnarray}
For brevity, we have introduced  $l_{1234}=\{l_{1},l_{2},l_{3},l_{4}\}$.
Thus we finally obtain
\begin{equation}
\label{W68}
L^{(e)}_{fi}(\tilde{\mathbf{q}})^{\beta \alpha}  =  \sum_{\kappa_{f},\bar{l},l} i^{l-\bar{l}} [\bar{l}l]^{2} \left[
\bar{v}_{e}^{2} L^{\beta \alpha}_{v_{e}v_{e}}+\bar{a}_{e}^{2} L^{\beta \alpha }_{a_{e}a_{e}}
+ \bar{v}_{e}\bar{a}_{e} (L^{\beta \alpha}_{v_{e}a_{e}} +L^{\beta \alpha}_{a_{e}v_{e}})\right],
\end{equation} 
\end{widetext}
where $\tilde{\mathbf{q}} \equiv (0,0,q)$. Explicit expressions for 
$L^{\beta \alpha}_{v_{e}v_{e}}$, etc are given in Appendix B.

The cross section (\ref{W42}), summed over all possible initial and final electron states, is
\begin{eqnarray}
\label{W69}
d \sigma^{(\nu)} & = & \frac{G_{\mathrm{F}}^{2}}{(2\pi)^{2} 32 m_{e}E_{\nu_{i}}} 
\int \delta(E_{fi}^{(\nu)})  E_{\nu_{f}}
d E_{\nu_{f}}d \Omega_{\nu_{f}} d E_{f}  \nonumber  \\
&& \times 
L_{fi}(\tilde{\mathbf{q}},p_{\nu_{i}},p_{\nu_{f}}),
\end{eqnarray}
where
\begin{eqnarray}
\label{W69b}
L_{fi}(\tilde{\mathbf{q}},p_{\nu_{i}},p_{\nu_{f}}) & = &
\mathrm{Re}[L^{(e)}_{fi}(\tilde{\mathbf{q}})^{\beta \alpha}]
\mathrm{Re}[ L^{(\nu)}(p_{\nu_{i}},p_{\nu_{f}})_{\beta \alpha} ]  \nonumber  \\
&& -
\mathrm{Im}[L^{(e)}_{fi}(\tilde{\mathbf{q}})^{\beta \alpha}]
\mathrm{Im}[L^{(\nu)}(p_{\nu_{i}},p_{\nu_{f}})_{\beta \alpha}], \nonumber  \\
\end{eqnarray}
and we have used, for final electron states normalized according to (\ref{B9}), $d\rho_{f}^{(e)}=dE_{f}/m_{e}$.
Introducing
\begin{eqnarray}
\label{W69a}
R^{\beta \alpha }(\tilde{\mathbf{q}}) &= &\int dE_{f} \delta(E_{fi}^{(\nu)})  
L^{(e)}_{fi}(\tilde{\mathbf{q}}) ^{\beta \alpha}  \nonumber  \\
& = & \sum_{\kappa_{f},\mu_{f},\mu_{i}} \int dE_{f}\delta(E_{fi}^{(\nu)})  
N_{fi}^{\beta}(\tilde{\mathbf{q}})^{*} N_{fi}^{\alpha}(\tilde{\mathbf{q}}),
\end{eqnarray}
then the cross section has the form used by~\cite{Chen2014,Chen2015} with their 
response functions (\ref{W09}) corresponding to (\ref{W69a}).

Of particular interest is the energy spectrum of the ionization electrons
\begin{equation}
\label{W70}
\frac{d \sigma^{(\nu)} }{d E_{f}} =  \frac{G_{\mathrm{F}}^{2}}{(2\pi)^{2}}
\frac{E_{\nu_{f}}}{32 m_{e} E_{\nu_{i}}}  \int  d \Omega_{\nu_{f}}
L_{fi}(\tilde{\mathbf{q}},p_{\nu_{i}},p_{\nu_{f}})
\end{equation}
where, in (\ref{W70}), it is understood that $E_{\nu_{f}}=E_{i}+E_{\nu_{i}}-E_{f}$.
As the coordinate system has been chosen such that $\mathbf{p}_{\nu_{i}}$, $\mathbf{p}_{\nu_{f}}$ and 
$\mathbf{q}$ lie in the $x-z$ plane with $\mathbf{q}$ along the $Oz$ axis, the 
integration over $\Omega_{\nu_{f}}$ becomes $-2 \pi \int d(\cos \theta )$ where $\theta $ is the angle
between $\mathbf{p_{\nu_{i}}}$ and $\mathbf{p}_{\nu_{f}}$. Noting that
$q^{2}=E_{\nu_{i}}^{2}+E_{\nu_{f}}^{2}-2E_{\nu_{i}}E_{\nu_{f}} \cos \theta $, then
$-d(\cos \theta )=dq^{2}/(2E_{\nu_{i}}E_{\nu_{f}})$, and our final expression for the energy
spectrum is
\begin{equation}
\label{W70a}
\frac{d \sigma^{(\nu)} }{d E_{f}} =  \frac{G_{\mathrm{F}}^{2}}{8\pi}
\frac{1}{16 m_{e} E_{\nu_{i}}^{2}}  \int  d q^{2} 
L_{fi}(\tilde{\mathbf{q}},p_{\nu_{i}},p_{\nu_{f}}).
\end{equation}

\section{Radial matrix elements}

The radial integrals (\ref{W53}) involve the Dirac radial functions $g_{\kappa,E}(r)$ 
and $f_{\kappa,E}(r)$ for the initial bound electron and the final continuum electron. For a 
Coulombic potential $V(r)=-\alpha Z/r$, the radial Dirac equations (\ref{W48}) have analytic 
solutions~\cite{Rose1961} in terms of confluent hypergeometric functions ${}_{1}F_{1}(a,c,z)$.

In this study we consider the scattering by electrons in the ground states of H, He and Ne. As
these systems involve only K- and L- shell electrons, we can use the simplified expressions
\begin{eqnarray}
\label{W71}
\left ( \begin{array}{cc}
g_{\kappa_{i},E_{i}}(r) \\
f_{\kappa_{i},E_{i}}(r) 
\end{array}
\right)
& = &
N_{i} \left( \begin{array}{cc}
\sqrt{m_{e}+E_{i}} \\
- \sqrt{m_{e}-E_{i}} 
\end{array}
\right) (2 \lambda_{i} r)^{\gamma_{i}}e^{-\lambda_{i}r} 
\nonumber \\
&& \times \left[ \left( \begin{array} {cc}
c_{0} \\
a_{0} 
\end{array} \right) + \left( \begin{array} {cc}
c_{1} \\
a_{1}
\end{array}  \right) \lambda_{i} r \right]
\end{eqnarray}
where
\begin{equation}
\label{W72}
\lambda_{i} \equiv \sqrt{m_{e}^{2}-E_{i}^{2}}, \quad \gamma_{i} \equiv \sqrt{\kappa_{i}^{2}-(\alpha Z)^{2}}
\end{equation}
and the initial state energy $E_{i}$ is
\begin{equation}
\label{W73}
E_{n,\kappa_{i}}= m_{e}\left[1 +\left(\frac{\alpha Z}{n-|\kappa_{i}|+\gamma_{i}} \right)^{2} \right]^{-1/2}.
\end{equation}
The dimensionless coefficients $(c_{0,1},a_{0,1}, N_{i})$ for the K-shell ($n=1, \kappa_{i}=-1$), 
L$_{\mathrm{I}}$ - subshell 
($n=2, \kappa_{i}=-1$), L$_{\mathrm{II}}$ -subshell ($n=2, \kappa_{i}=+1$) and 
L$_{\mathrm{III}}$ -subshell ($n=2,\kappa_{i}=-2$) are tabulated in~\cite{Rose1961}.
(Note that \cite{Rose1961} uses relativistic units $\hbar = c= m_{e} =1$.)
Since $\alpha Z \ll 1$, the initial state binding energy $\epsilon_{i} \equiv m_{e}-E_{i}$ can be 
approximated as $(\alpha Z)^{2}m_{e}/2$ for K-shell electrons and $(\alpha Z)^{2}m_{e}/8 $ 
for L-shell electrons.
The screening effects of the electrons in the filled K- shell (for He and Ne), and 
L-subshells (for Ne) are represented by an effective nuclear charge $Z_{\mathrm{eff}}=Z-s_{i}$,
a procedure that should be reasonable for small principal quantum number $n$ and small 
$n-l$~\cite{Bethe1957}. These screening constants $s_{i}$, taken from the 
fits~\cite{Thomas1997} of Dirac single electron eigenfunctions to empirical binding energies, are
0.656 (K-shell), 2.016 (L$_{\mathrm{I}}$ -subshell), 6.254 (L$_{\mathrm{II}}$ -subshell)
and 7.482 (L$_{\mathrm{III}}$ -subshell).

The final electron continuum states, energy normalized according to (\ref{B9}), are
\begin{eqnarray}
\label{W74}
\left ( \begin{array}{cc}
g_{\kappa_{f},E_{f}}(r) \\
f_{\kappa_{f},E_{f}}(r) 
\end{array}
\right)
& = &
N_{f} \left( \begin{array}{cc}
\sqrt{E_{f}+m_{e}} \\
i \sqrt{E_{f}-m_{e}} 
\end{array}
\right) (2 p_{f} r)^{\gamma_{f}}[ (\gamma_{f}+iy)
\nonumber  \\ 
&& \times e^{-ip_{f}r+i\eta} {}_{1}F_{1}(a,c, 2ip_{f}r) \pm c.c],
\end{eqnarray}
where 
\begin{equation}
\label{W75}
\gamma_{f}\equiv \sqrt{\kappa_{f}^{2}-(\alpha Z)^{2}}, \quad p_{f} \equiv \sqrt{E_{f}^{2}-m_{e}^{2}},
\quad y \equiv \frac{\alpha Z E_{f}}{p_{f}},
\end{equation}
and
\begin{equation}
\label{W76}
e^{2i\eta} = -\frac{\kappa_{f}-iym_{e}/E_{f}}{\gamma_{f}+iy}.
\end{equation}
The dimensionless normalization constant is
\begin{equation}
\label{W77}
N_{f}=\frac{e^{\pi y/2}}{2} \sqrt{\frac{m_{e}}{\pi p_{f}}} \; \frac{|\Gamma(\gamma_{f}+iy)|}
{\Gamma (2 \gamma_{f}+1)}.
\end{equation}
The parameters in the hypergeometric function are $a = \gamma_{f}+1+iy$ and $c=2 \gamma_{f}+1$. 
Since $a$ and $z\equiv 2ip_{f}r$ are complex, the computation of ${}_{1}F_{1}(a,c,z)$ involves the 
summation of a slowly convergent complex series for each required value of $z$. Consequently
we choose to integrate the Dirac equation directly.
For each shell and subshell calculation, the continuum state electrons are assumed to move in the 
same potential as the bound state electrons~\cite{Schofield1973}. 

Details of the computation of the continuum radial functions and the radial integrals 
(\ref{W53}) are given in Appendix C.

\section{Evaluation of cross sections}

The cross section involves the contraction of the electron tensor $L^{(e)}_{fi}(\mathbf{q}) ^{\beta \alpha}$, 
given by (\ref{W68}),  with the neutrino tensor $L^{(\nu )}_{\beta \alpha}(p_{\nu_{i}},p_{\nu_{f}})$ 
given by (\ref{W44}).
From (\ref{W67}), we need only the diagonal elements $(\beta, \beta )$
and the off-diagonal elements $(\beta, \alpha)=(1,2),(2,1),(0,3),(3,0)$ 
of the two tensors. 

The summation over $l$ in the electron tensor is constrained by the conditions
$\bar{\Delta} (l_{3},l_{4},l)$ where $l_{3}=(l_{f},l_{f}^{\prime})$ and 
$l_{4}=(l_{i},l_{i}^{\prime})$. Here 
$\bar{\Delta} (a,b,c)$ implies $|a-b| \le c \le a+b$, together with $a+b+c = \mathrm{even \;integer}$.
Similar constraints apply to the summation over $\bar{l}$. As $l_{i}$ has the values 0 or 1, 
and $l_{i}^{\prime}$ the values 0, 1 or 2, the number of terms in these summations is quite small.
However, the summation over $\kappa_{f}$ is unconstrained, with convergence coming from the 
decreasing overlap between the initial bound and final continuum electron eigenfunctions
with increasing $\kappa_{f}$  in the radial integrals. 

For $\kappa_{i}=-1$, the special case
\begin{equation}
\label{W80}
A^{(I,I)}_{\bar{l}l}(l_{f},0,l_{f},0) = 
A^{(\lambda,\lambda)}_{\bar{l}l}(l_{f},0,l_{f},0)
=\delta_{\bar{l},l} \delta_{l,l_{f}} \frac{2j_{f}+1}{(2l_{f}+1)^{2}}
\end{equation}
can be used to check the evaluation of the $A$ coefficients. A 
similar result holds for $\kappa_{i}=1$ with $l_{f}$ replaced by $l_{f}^{\prime}$.

The required elements of the neutrino tensor are calculated from (\ref{W44}) 
where, with our choice of coordinate system, 
$\mathbf{p}_{\nu_{i}}=(E_{\nu_{i}} \sin (\gamma-\theta) ,0, E_{\nu_{i}}\cos (\gamma -\theta) )$
and
$\mathbf{p}_{\nu_{f}}=(E_{\nu_{f}} \sin \gamma ,0, E_{\nu_{f}}\cos \gamma )$.
Here, $\gamma$ is the angle between $\mathbf{q}$ and $\mathbf{p}_{\nu_{f}}$
and is related to the scattering angle $\theta$ via
\begin{equation}
\label{W70c}
\tan \gamma = \frac{\sin \theta }{\cos \theta -E_{\nu_{f}}/ E_{\nu_{i}} }.
\end{equation}
These elements can be expressed~\cite{Chen2014} in terms of the energy transfer 
$T \equiv E_{\nu_{i}}-E_{\nu_{f}}$ and the quantity $Q^{2} \equiv q^{2}-T^{2}$, that is, $Q^{2}=-t >0$.
Explicitly,
\begin{eqnarray}
\label{W82}
L^{(\nu)}{}^{0,0} & = & 16 E_{\nu_{i}}E_{\nu_{f}} \cos^{2} (\frac{\theta }{2}), \nonumber  \\
L^{(\nu)}{}^{1,1} & = & 16 E_{\nu_{i}}E_{\nu_{f}} \cos^{2} (\frac{\theta }{2}) 
\left[\tan^{2} (\frac{\theta}{2})+ \frac{Q^{2}}{q^{2}} \right], \nonumber \\
L^{(\nu)}{}^{2,2} & = & 16 E_{\nu_{i}}E_{\nu_{f}} \cos^{2} (\frac{\theta }{2}) 
\tan^{2} (\frac{\theta}{2}), \nonumber \\
L^{(\nu)}{}^{3,3} & = & 16 E_{\nu_{i}}E_{\nu_{f}} \cos^{2} (\frac{\theta }{2}) 
\frac{T^{2}}{q^{2}}, \nonumber  \\
L^{(\nu)}{}^{1,2} & = & -  L^{(\nu)}{}^{2,1} =- 16 i E_{\nu_{i}}E_{\nu_{f}} \cos^{2} 
(\frac{\theta }{2}) \tan (\frac{\theta }{2}) \nonumber  \\
& & \times \sqrt{\tan^{2} (\frac{\theta}{2})+ \frac{Q^{2}}{q^{2}}}, \nonumber  \\
L^{(\nu)}{}^{0,3} & = & L^{(\nu)}{}^{3,0} = 16 E_{\nu_{i}}E_{\nu_{f}} 
\cos^{2}(\frac{\theta }{2}) \;\frac{T}{q}. 
\end{eqnarray}

For antineutrino scattering, $ L^{(\bar{\nu})}{}^{\beta, \alpha}  = (L^{(\nu)}{}^{\beta ,\alpha})^{*}$. 
The difference between $\nu_{e}$ and $\bar{\nu}_{e}$ scattering  therefore arises solely from the $(1,2)$ and $(2,1)$
components.

\section{Results and discussion}

The energy spectra $d \sigma/dE_{f}$ of the ionization electrons produced in low energy scattering
of electron neutrinos and antineutrinos by atomic electrons have been calculated as a function of the 
electron kinetic energy $\epsilon_{f} = E_{f}-m_{e}$.
Results are obtained for scattering of 5, 10, 20, and 30 keV neutrino energies
by the ground state systems H($1s$), He($1s^{2}$), 
and Ne($1s^{2}2s^{2}2p^{6}$) where, for He and Ne, the electrons are considered as independent scattering
centers.

The energy spectra are compared to that for scattering from free electrons, for which,
in the laboratory frame, $E_{i}= m_{e}, \;\mathbf{p}_{e_{i}}=0$ and the kinematic variables simplify to
\begin{eqnarray}
\label{W101}
s&=& m_{e}(m_{e}+2 E_{\nu_{i}}), \nonumber  \\
u&=& m_{e}(m_{e}-2 E_{\nu_{f}}), \nonumber  \\
t&=& 2m_{e}(m_{e} -E_{f}).
\end{eqnarray}
Setting $\tilde{m}=m_{e}$ in (\ref{W01}), the energy spectrum of the scattered electron 
for $\nu_{e}$ scattering is then~\cite{Gounaris2002}
\begin{eqnarray}
\label{W102}
\left(\frac{d \sigma^{(\nu)}}{d E_{f}}\right)_{(\mathrm{Free})}  & = & 
\frac{ G_{F}^{2} m_{e}}{8 \pi E_{\nu_{i}}^{2}} \left\{
(\bar{v}_{e}-\bar{a}_{e})^{2} E_{\nu_{i}}^{2} \right.
\nonumber  \\
&& + (\bar{v}_{e}+\bar{a}_{e})^{2}(E_{\nu_{i}}+m_{e}-E_{f})^{2}
\nonumber  \\
&& \left. + m_{e}(\bar{v}_{e}^{2}-\bar{a}_{e}^{2}) (m_{e}-E_{f}) \right\},
\end{eqnarray}
where $m_{e} \le E_{f} \le m_{e}+ \epsilon_{f}^{\mathrm{max}}$ and the maximum kinetic energy is
\begin{equation}
\label{W103}
\epsilon_{f}^{\mathrm{max}} = \frac{2 E_{\nu_{i}}^{2}}{m_{e}+2E_{\nu_{i}}}.
\end{equation}
For low energy transfers $T \ll E_{\nu_{i}}$, 
\begin{eqnarray}
\label{W109}
\left(\frac{d \sigma^{(\nu)}}{d E_{f}}\right)_{(\mathrm{Free})} & = &
%\frac{d \sigma^{(\nu)}_{(\mathrm{Free})}}{d E_{f}} & = & 
\frac{ G_{F}^{2} m_{e}}{2 \pi }(1+4 \sin^{2}\theta_{W} + 8 \sin^{4} \theta_{W})
\nonumber  \\
&& \times \left[ 1+O\left(\frac{T}{E_{\nu_{i}}}\right)\right].
\end{eqnarray}

The total cross section for scattering off free electrons is~\cite{Gounaris2002}
\begin{eqnarray}
\label{W102a}
\sigma^{(\nu)}_{(\mathrm{Free})}  & = & 
\frac{ G_{F}^{2} m_{e}E_{\nu_{i}}}{8 \pi } \left[
(\bar{v}_{e}-\bar{a}_{e})^{2} \frac{2 E_{\nu_{i}}}{m_{e}+2 E_{\nu_{i}}} \right.
\nonumber  \\
&& + \frac{1}{3}(\bar{v}_{e}+\bar{a}_{e})^{2}\left\{1-\frac{m_{e}^{3}}{(m_{e}+2 E_{\nu_{i}})^{3}}\right\}
\nonumber  \\
&& \left. - (\bar{v}_{e}^{2}-\bar{a}_{e}^{2}) \frac{2 m_{e} E_{\nu_{i}}}{(m_{e}+2 E_{\nu_{i}})^{2} }\right].
\end{eqnarray}

For $\bar{\nu}_{e}$ scattering, the interchange $s \leftrightarrow u$ in (\ref{W01})
is equivalent to $\bar{a}_{e} \leftrightarrow -\bar{a}_{e}$ in (\ref{W102}) and
(\ref{W102a}). The low $T$ limit is unaltered.

Energy spectra and  total cross sections for $\nu_{e}$ and $\bar{\nu}_{e}$ scattering by free 
electrons are given in Table I. 

Results for $\nu_{e}$ ($\bar{\nu}_{e}$) scattering by H are given in Table II (Table III), 
by He in Table IV (Table V), and by Ne in Table VI (Table VII), respectively. 
The energy spectra and cross sections are expressed as ratios 
\begin{equation}
\label{W100a}
R^{(\nu)}(E_{f}) = \frac{d\sigma ^{(\nu)}/dE_{f}}{Z(d\sigma ^{(\nu)}/dE_{f})_{(\mathrm{Free})}},
\end{equation}
and $\sigma^{(\nu)}/Z\sigma^{(\nu)}_{(\mathrm{Free})}$,
to the corresponding quantities for scattering by $Z$ free electrons. 
Also listed are
results for the case where the final continuum electron is treated as 
free, for which the  radial eigenfunctions normalized according to (\ref{B9}) are
\begin{equation}
\label{W100}
\left ( \begin{array}{cc}
g_{\kappa_{f},E_{f}}(r) \\
f_{\kappa_{f},E_{f}}(r) 
\end{array}
\right)
=
\sqrt{\frac{p_{f} m_{e}}{\pi}} \left( \begin{array}{cc}
\sqrt{E_{f}+m_{e}} \;r j_{l_{f}}(p_{f}r) \\
S_{\kappa_{f}} \sqrt{E_{f}-m_{e}} \;r j_{l_{f}^{\prime}}(p_{f}r)
\end{array}
\right), 
\end{equation}
where $S_{\kappa_{f}} \equiv \kappa_{f}/|\kappa_{f}|$.

\begin{table*}
\caption{\label{tab:free}Energy spectra $(d\sigma^{(\nu)}/dE_{f}) _{(\mathrm{Free})}$ 
of electrons resulting from scattering of incident neutrinos of energy $E_{\nu_{i}}$ by free electrons. 
The results, in units of $10^{-14}$ GeV$^{-3}$, are given as a function of the kinetic 
energy $\epsilon_{f}$ of the electron.  
Also shown are the integrated spectra $\sigma^{(\nu)}_{(\mathrm{Free})}$  
in units of $10^{-20}$ GeV$^{-2}$.
Results for scattering of antineutrinos are given in parentheses.}
\begin{ruledtabular}
\begin{tabular}{lllll}
$\epsilon_{f}/\epsilon_{f}^{\mathrm{max}}$ & $E_{\nu_{i}}=5 $ (keV)&  $E_{\nu_{i}}=10$ (keV) &$ E_{\nu_{i}}=20$ (keV) & $E_{\nu_{i}}=30$ (keV) \\
\hline
0.0  & 2.6029(2.6029) & 2.6029(2.6029) & 2.6029(2.6029) & 2.6029 (2.6029) \\
0.1  & 2.4553(2.4471) & 2.4571(2.4411) & 2.4607(2.4299) & 2.4641 (2.4195) \\
0.2  & 2.3076(2.2912) & 2.3114(2.2794) & 2.3185(2.2571) & 2.3252 (2.2367) \\
0.3  & 2.1599(2.1354) & 2.1656(2.1177) & 2.1764(2.0846) & 2.1865 (2.0543) \\
0.4  & 2.0122(1.9796) & 2.0198(1.9561) & 2.0342(1.9124) & 2.0478 (1.8725)   \\
0.5  & 1.8645(1.8238) & 1.8740(1.7946) & 1.8921(1.7403) & 1.9091 (1.6912)   \\
0.6  & 1.7168(1.6681) & 1.7283(1.6331) & 1.7500(1.5686) & 1.7705 (1.5104)   \\
0.7  & 1.5692(1.5123) & 1.5825(1.4717) & 1.6080(1.3970) & 1.6319 (1.3301)   \\
0.8  & 1.4215(1.3566) & 1.4368(1.3104) & 1.4660(1.2258) & 1.4934 (1.1504)   \\
0.9  & 1.2738(1.2009) & 1.2910(1.1491) & 1.3239(1.0547) & 1.3550 (0.9712)   \\
1.0  & 1.1261(1.0452) & 1.1453(0.9879) & 1.1820(0.8840) & 1.2166 (0.7925)   \\
$\epsilon_{f}^{\mathrm{max}}$ (eV) & 95.969 & 376.65 & 1451.9 & 3152.4  \\
$\sigma^{(\nu)}_{(\mathrm{Free})}$ &  0.31980(0.31590) &  1.2483(1.2185) & 4.7624(4.5434) & 10.241(9.5600) \\
\end{tabular}
\end{ruledtabular}
\end{table*}

\begin{table*}
\caption{\label{tab:nu-H}Energy spectra $d\sigma^{(\nu)} /dE_{f}$  of the ionization electrons from 
scattering of incident neutrinos of energy $E_{\nu_{i}}$ by hydrogen. The results are 
expressed as ratios to the spectra $(d\sigma^{(\nu)}/dE_{f}) _{(\mathrm{Free})}$  
for scattering by free electrons and are given as a function of the kinetic energy $\epsilon_{f}$ of the electron . 
Results are shown for a Coulombic final electron and, in parentheses, for a free final electron. 
Also shown are the integrated spectra $\sigma^{(\nu)}$ expressed as a ratio to the integrated spectra
$\sigma^{(\nu)}_{(\mathrm{Free})}$ for a free electron.}
\begin{ruledtabular}
\begin{tabular}{lllll}
$\epsilon_{f}/\epsilon_{f}^{\mathrm{max}}$ & $E_{\nu_{i}}=5 $ (keV)&  $E_{\nu_{i}}=10$ (keV) &$ E_{\nu_{i}}=20$ (keV) & $E_{\nu_{i}}=30$ (keV) \\
\hline
0.1  & 0.7780(0.8467) & 0.9472(0.9776) & 0.9880(0.9970) & 0.9952 (0.9991) \\
0.2  & 0.7680(0.9169) & 0.9465(0.9877) & 0.9877(0.9976) & 0.9953 (0.9994) \\
0.3  & 0.7482(0.9345) & 0.9427(0.9890) & 0.9870(0.9976) & 0.9952 (0.9995) \\
0.4  & 0.7192(0.9354) & 0.9368(0.9883) & 0.9861(0.9976) & 0.9945 (0.9991)   \\
0.5  & 0.6795(0.9254) & 0.9267(0.9857) & 0.9849(0.9974) & 0.9930 (0.9981)   \\
0.6  & 0.6273(0.9033) & 0.9074(0.9782) & 0.9828(0.9968) & 0.9906 (0.9961)   \\
0.7  & 0.5624(0.8664) & 0.8669(0.9574) & 0.9773(0.9942) & 0.9868 (0.9931)   \\
0.8  & 0.4872(0.8125) & 0.7819(0.9022) & 0.9545(0.9796) & 0.9780 (0.9864)   \\
0.9  & 0.4075(0.7428) & 0.6286(0.7797) & 0.8376(0.8877) & 0.9183 (0.9383)   \\
1.0  & 0.3308(0.6630) & 0.4256(0.5840) & 0.4629(0.5367) & 0.4643 (0.5122)   \\
$\epsilon_{f}^{\mathrm{max}}$ (eV) & 95.969 & 376.65 & 1451.9 & 3152.4  \\
$\sigma^{(\nu)}/\sigma^{(\nu)}_{(\mathrm{Free})}$ &  0.6626(0.8696) &  0.8811(0.9472) & 0.9568(0.9741) & 0.9729(0.9802) \\
\end{tabular}
\end{ruledtabular}
\end{table*}
\begin{table*}
\caption{\label{tab:nubar-H}Energy spectra $d\sigma^{(\bar{\nu})} /dE_{f}$  of the ionization electrons from 
scattering of incident antineutrinos of energy $E_{\nu_{i}}$ by hydrogen. The results are 
expressed as ratios to the spectrum  $(d\sigma^{(\bar{\nu})}/dE_{f}) _{(\mathrm{Free})}$ 
for scattering by free electrons and are given as a function of the kinetic energy $\epsilon_{f}$ of the electron . 
Results are shown for a Coulombic final electron and, in parentheses, for a free final electron. 
Also shown are the integrated spectra $\sigma^{(\bar{\nu})}$ expressed as a ratio to the integrated spectra
$\sigma^{(\bar{\nu})}_{(\mathrm{Free})}$ for a free electron.}
 \begin{ruledtabular}
\begin{tabular}{lllll}
$\epsilon_{f}/\epsilon_{f}^{\mathrm{max}}$ & $E_{\nu_{i}}=5 $ (keV)&  $E_{\nu_{i}}=10$ (keV) &$ E_{\nu_{i}}=20$ (keV)& $E_{\nu_{i}}=30$ (keV) \\
\hline
0.1  & 0.7690(0.8448) & 0.9415(0.9768) & 0.9855(0.9969) & 0.9938 (0.9993)  \\
0.2  & 0.7587(0.9148) & 0.9401(0.9868) & 0.9850(0.9977) & 0.9940 (0.9999)  \\
0.3  & 0.7389(0.9324) & 0.9356(0.9879) & 0.9840(0.9979) & 0.9938 (1.000)  \\
0.4  & 0.7105(0.9336) & 0.9287(0.9870) & 0.9827(0.9978) & 0.9930 (1.000)  \\
0.5  & 0.6720(0.9244) & 0.9178(0.9841) & 0.9807(0.9975) & 0.9912 (0.9992)  \\
0.6  & 0.6219(0.9043) & 0.8985(0.9768) & 0.9775(0.9965) & 0.9882 (0.9972)  \\
0.7  & 0.5599(0.8706) & 0.8599(0.9576) & 0.9578(0.9934) & 0.9833 (0.9937)  \\
0.8  & 0.4882(0.8213) & 0.7804(0.9074) & 0.9485(0.9795) & 0.9731 (0.9862)  \\
0.9  & 0.4123(0.7577) & 0.6361(0.7943) & 0.8416(0.8966) & 0.9179 (0.9422)  \\
1.0  & 0.3393(0.6852) & 0.4415(0.6093) & 0.4844(0.5693) & 0.4880 (0.5399)  \\
$\epsilon_{f}^{\mathrm{max}}$ (eV) & 95.969 & 376.65 & 1451.9 & 3152.4  \\
$\sigma^{(\bar{\nu})}/\sigma^{(\bar{\nu})}_{(\mathrm{Free})}$ &  0.6590(0.8717) & 0.8794(0.9505) & 0.9576(0.9779)  & 0.9749(0.9843)  \\
\end{tabular}
\end{ruledtabular}
\end{table*}

\begin{table*}
\caption{\label{tab:nu-He}Energy spectra $d\sigma^{(\nu)} /dE_{f}$  of the ionization electrons from 
scattering of incident neutrinos of energy $E_{\nu_{i}}$ by helium. The results are 
expressed as ratios to the spectra $2(d\sigma^{(\nu)}/dE_{f}) _{(\mathrm{Free})}$  
for scattering by two free electrons and are given as a function of the kinetic energy $\epsilon_{f}$ of the electron . 
Results are shown for a Coulombic final electron and, in parentheses, for a free final electron. 
Also shown are the integrated spectra $\sigma^{(\nu)}$ expressed as a ratio to the integrated spectra
$2\sigma^{(\nu)}_{(\mathrm{Free})}$ for two free electrons.}
\begin{ruledtabular}
\begin{tabular}{lllll}
$\epsilon_{f}/\epsilon_{f}^{\mathrm{max}}$ & $E_{\nu_{i}}=5 $ (keV)&  $E_{\nu_{i}}=10$ (keV)&$ E_{\nu_{i}}=20$ (keV) & $E_{\nu_{i}}=30$  (keV)\\
%& R^{(\mathrm{free})}(\epsilon_{f}) & R^{(\mathrm{Coul})}(\epsilon_{f})  \\
\hline
0.1  & 0.6179(0.7209) & 0.9020(0.9444) & 0.9778(0.9929) & 0.9910 (0.9977) \\
0.2  & 0.5983(0.8258) & 0.9000(0.9719) & 0.9774(0.9951) & 0.9910 (0.9983) \\
0.3  & 0.5705(0.8607) & 0.8920(0.9766) & 0.9761(0.9954) & 0.9907 (0.9986) \\
0.4  & 0.5358(0.8692) & 0.8787(0.9757) & 0.9743(0.9952) & 0.9902 (0.9987)   \\
0.5  & 0.4952(0.8626) & 0.8568(0.9697) & 0.9715(0.9947) & 0.9894 (0.9986)   \\
0.6  & 0.4498(0.8444) & 0.8200(0.9549) & 0.9659(0.9929) & 0.9879 (0.9982)   \\
0.7  & 0.4015(0.8164) & 0.7585(0.9222) & 0.9512(0.9857) & 0.9842 (0.9964)   \\
0.8  & 0.3528(0.7801) & 0.6627(0.8579) & 0.9031(0.9549) & 0.9680 (0.9861)   \\
0.9  & 0.3062(0.7385) & 0.5331(0.7511) & 0.7513(0.8359) & 0.8671 (0.9068)   \\
1.0  & 0.2641(0.6953) & 0.3904(0.6088) & 0.4472(0.5492) & 0.4607 (0.5268)   \\
$\epsilon_{f}^{\mathrm{max}}$ (eV) & 95.969 & 376.65 & 1451.9 & 3152.4  \\
$\sigma^{(\nu)}/\sigma^{(\nu)}_{(\mathrm{Free})}$ &  0.5019(0.7988) &  0.8129(0.9235) & 0.9343(0.9644) & 0.9635(0.9769) \\
\end{tabular}
\end{ruledtabular}
\end{table*}
\begin{table*}
\caption{\label{tab:nubar-He}Energy spectra $d\sigma^{(\bar{\nu})} /dE_{f}$  of the ionization electrons from 
scattering of incident antineutrinos of energy $E_{\nu_{i}}$ by helium. The results are 
expressed as ratios to the spectrum  $2(d\sigma^{(\bar{\nu})}/dE_{f}) _{(\mathrm{Free})}$ 
for scattering by two free electrons and are given as a function of the kinetic energy $\epsilon_{f}$ of the electron . 
Results are shown for a Coulombic final electron and, in parentheses, for a free final electron. 
Also shown are the integrated spectra $\sigma^{(\bar{\nu})}$ expressed as a ratio to the integrated spectra
$2\sigma^{(\bar{\nu})}_{(\mathrm{Free})}$ for two free electrons.}
 \begin{ruledtabular}
\begin{tabular}{lllll}
$\epsilon_{f}/\epsilon_{f}^{\mathrm{max}}$ & $E_{\nu_{i}}=5 $ (keV)&  $E_{\nu_{i}}=10$ (keV)&$ E_{\nu_{i}}=20$ (keV)& $E_{\nu_{i}}=30$ (keV) \\
\hline
0.1  & 0.6070(0.7180) & 0.8919(0.9427) & 0.9729(0.9925) & 0.9880 (0.9977)  \\
0.2  & 0.5877(0.8226) & 0.8889(0.9699) & 0.9720(0.9949) & 0.9879 (0.9987)  \\
0.3  & 0.5608(0.8579) & 0.8800(0.9744) & 0.9700(0.9951) & 0.9874 (0.9991)  \\
0.4  & 0.5274(0.8673) & 0.8659(0.9732) & 0.9672(0.9947) & 0.9864 (0.9994)  \\
0.5  & 0.4885(0.8624) & 0.8441(0.9674) & 0.9630(0.9938) & 0.9848 (0.9992)  \\
0.6  & 0.4454(0.8469) & 0.8090(0.9539) & 0.9560(0.9914) & 0.9821 (0.9984)  \\
0.7  & 0.3997(0.8227) & 0.7517(0.9249) & 0.9408(0.9842) & 0.9766 (0.9959)  \\
0.8  & 0.3538(0.7917) & 0.6629(0.8678) & 0.8967(0.9571) & 0.9600 (0.9856)  \\
0.9  & 0.3102(0.7569) & 0.5423(0.7717) & 0.7597(0.8520) & 0.8705 (0.9167)  \\
1.0  & 0.2712(0.7224) & 0.4074(0.6413) & 0.4732(0.5848) & 0.4906 (0.5637)  \\
$\epsilon_{f}^{\mathrm{max}}$ (eV) & 95.969 & 376.65 & 1451.9 & 3152.4  \\
$\sigma^{(\bar{\nu})}/\sigma^{(\bar{\nu})}_{(\mathrm{Free})}$ &  0.4975(0.8008) & 0.8091(0.9272) & 0.9340(0.9696)  & 0.9649(0.9818)  \\
\end{tabular}
\end{ruledtabular}
\end{table*}
 
\begin{table*}
\caption{\label{tab:nu-Ne}Energy spectra $d\sigma^{(\nu)} /dE_{f}$  of the ionization electrons from 
scattering of incident neutrinos of energy $E_{\nu_{i}}$ by neon. The results are 
expressed as ratios to the spectra $10(d\sigma^{(\nu)}/dE_{f}) _{(\mathrm{Free})}$  
for scattering by 10 free electrons and are given as a function of the kinetic energy $\epsilon_{f}$ of the electron . 
Results are shown for a Coulombic final electron and, in parentheses, for a free final electron. 
Also shown are the integrated spectra $\sigma^{(\nu)}$ expressed as a ratio to the integrated spectra
$10\sigma^{(\nu)}_{(\mathrm{Free})}$ for 10 free electrons.}
\begin{ruledtabular}
\begin{tabular}{lllll}
$\epsilon_{f}/\epsilon_{f}^{\mathrm{max}}$ & $E_{\nu_{i}}=5 $ (keV)&  $E_{\nu_{i}}=10$ (keV) &$ E_{\nu_{i}}=20$ (keV) & $E_{\nu_{i}}=30$ (keV) \\
\hline
0.1  & 0.09921(0.2905) & 0.2847(0.4001) & 0.4694(0.5130) & 0.5697 (0.5901) \\
0.2  & 0.09333(0.2971) & 0.2783(0.4188) & 0.4789(0.5533) & 0.5793 (0.6365) \\
0.3  & 0.08777(0.2892) & 0.2685(0.4277) & 0.4818(0.5755) & 0.5798 (0.6561) \\
0.4  & 0.08329(0.2760) & 0.2554(0.4310) & 0.4798(0.5882) & 0.5757 (0.6641)   \\
0.5  & 0.08068(0.2594) & 0.2384(0.4283) & 0.4720(0.5951) & 0.5691 (0.6652)   \\
0.6  & 0.08027(0.2411) & 0.2164(0.4167) & 0.4548(0.5967) & 0.5597 (0.6617)   \\
0.7  & 0.08149(0.2244) & 0.1908(0.3922) & 0.4225(0.5885) & 0.5424 (0.6536)   \\
0.8  & 0.08286(0.2131) & 0.1696(0.3555) & 0.3641(0.5543) & 0.4997 (0.6314)   \\
0.9  & 0.08277(0.2100) & 0.1623(0.3235) & 0.2773(0.4672) & 0.3880 (0.5438)   \\
1.0  & 0.08036(0.2161) & 0.1578(0.3157) & 0.2334(0.3973) & 0.2720 (0.4043)   \\
$\epsilon_{f}^{\mathrm{max}}$ (eV) & 95.969 & 376.65 & 1451.9 & 3152.4  \\
$\sigma^{(\nu)}/\sigma^{(\nu)}_{(\mathrm{Free})}$ &  0.08732(0.2613) &  0.2383(0.3998) & 0.4375(0.5494) & 0.5395(0.6220) \\
\end{tabular}
\end{ruledtabular}
\end{table*}
\begin{table*}
\caption{\label{tab:nubar-Ne}Energy spectra $d\sigma^{(\bar{\nu})} /dE_{f}$  of the ionization electrons from 
scattering of incident antineutrinos of energy $E_{\nu_{i}}$ by neon. The results are 
expressed as ratios to the spectrum  $10(d\sigma^{(\bar{\nu})}/dE_{f}) _{(\mathrm{Free})}$ 
for scattering by 10 free electrons and are given as a function of the kinetic energy $\epsilon_{f}$ of the electron . 
Results are shown for a Coulombic final electron and, in parentheses, for a free final electron. 
Also shown are the integrated spectra $\sigma^{(\bar{\nu})}$ expressed as a ratio to the integrated spectra
$10\sigma^{(\bar{\nu})}_{(\mathrm{Free})}$ for 10 free electrons.}
 \begin{ruledtabular}
\begin{tabular}{lllll}
$\epsilon_{f}/\epsilon_{f}^{\mathrm{max}}$ & $E_{\nu_{i}}=5 $ (keV)&  $E_{\nu_{i}}=10$ (keV) &$ E_{\nu_{i}}=20$ (keV)& $E_{\nu_{i}}=30$ (keV) \\
\hline
0.1  & 0.09751(0.2881) & 0.2780(0.3981) & 0.4588(0.5100) & 0.5539 (0.5865)  \\
0.2  & 0.09183(0.2946) & 0.2721(0.4169) & 0.4686(0.5509) & 0.5651 (0.6338)  \\
0.3  & 0.08653(0.2870) & 0.2630(0.4260) & 0.4716(0.5747) & 0.5675 (0.6553)  \\
0.4  & 0.08230(0.2746) & 0.2507(0.4299) & 0.4700(0.5894) & 0.5656 (0.6661)  \\
0.5  & 0.07993(0.2592) & 0.2349(0.4284) & 0.4630(0.5987) & 0.5611 (0.6711)  \\
0.6  & 0.07975(0.2424) & 0.2145(0.4191) & 0.4479(0.6036) & 0.5538 (0.6724)  \\
0.7  & 0.08126(0.2274) & 0.1909(0.3982) & 0.4194(0.6005) & 0.5397 (0.6705)  \\
0.8  & 0.08310(0.2179) & 0.1718(0.3667) & 0.3669(0.5751) & 0.5036 (0.6577)  \\
0.9  & 0.08373(0.2171) & 0.1664(0.3405) & 0.2876(0.5015) & 0.4033 (0.5863)  \\
1.0  & 0.08234(0.2263) & 0.1650(0.3406) & 0.2509(0.4474) & 0.3138 (0.4671)  \\
$\epsilon_{f}^{\mathrm{max}}$ (eV) & 95.969 & 376.65 & 1451.9 & 3152.4  \\
$\sigma^{(\bar{\nu})}/\sigma^{(\bar{\nu})}_{(\mathrm{Free})}$ &  0.08669(0.2586) & 0.2366(0.4029) & 0.4351(0.5558)  & 0.5375(0.6309)  \\
\end{tabular}
\end{ruledtabular}
\end{table*}

\begin{figure}[ht]
\begin{center}
\includegraphics[width=0.95\columnwidth]{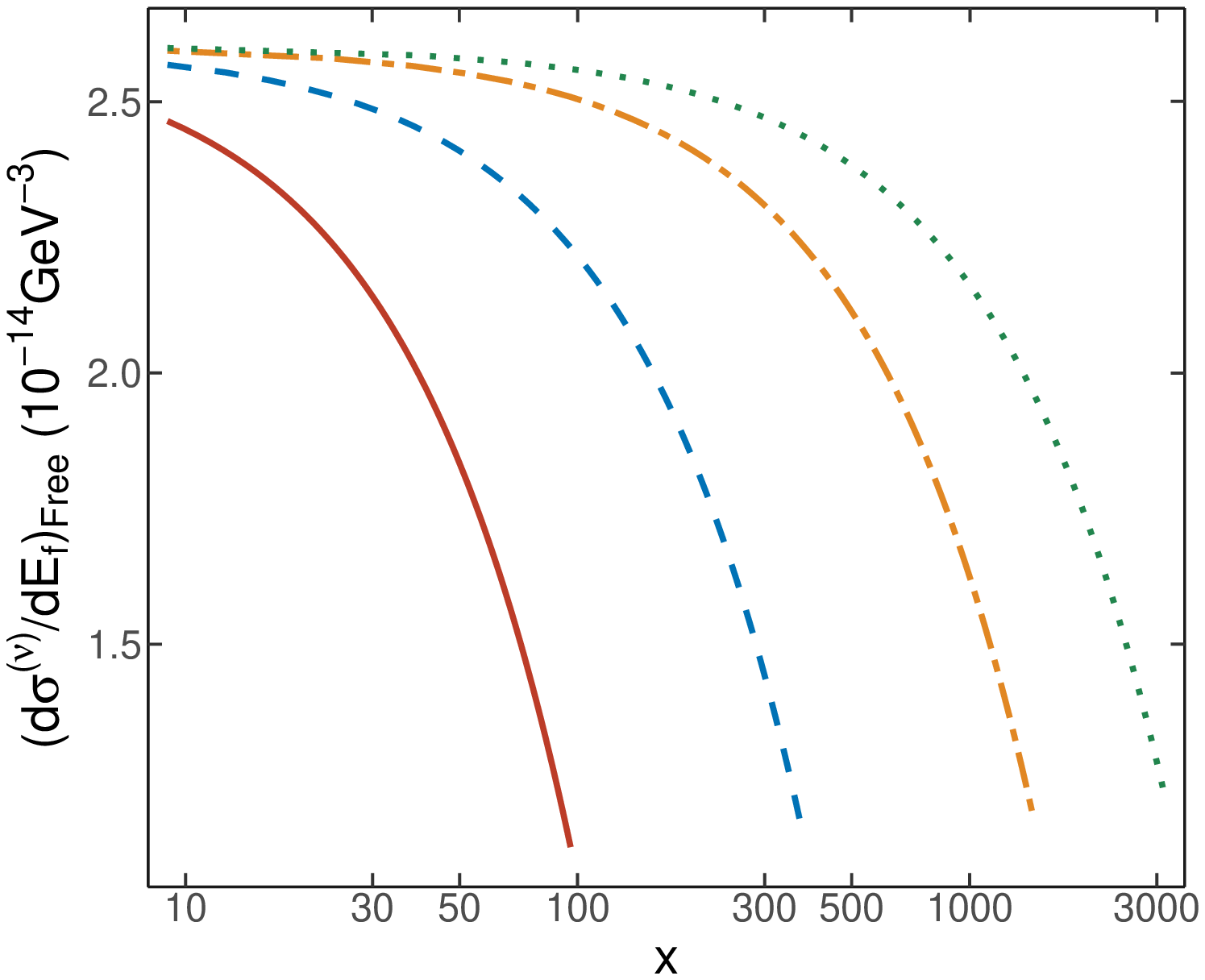}
\caption{\label{fig:free}Energy spectra  $(d\sigma^{(\nu)}/dE_{f}) _{(\mathrm{Free})}$ (Eqn (\ref{W102})), 
as a function of electron kinetic energy $\epsilon_{f}$,
of electrons resulting from scattering of neutrinos by free electrons. 
Results are shown for scattering of 5 keV (solid line), 10 keV (dashed line), 20 keV (dash-dotted line), 
and 30 keV (dotted line) incident neutrino energies. }
\end{center}
\end{figure}

\begin{figure}[ht]
\includegraphics[width=0.95\columnwidth]{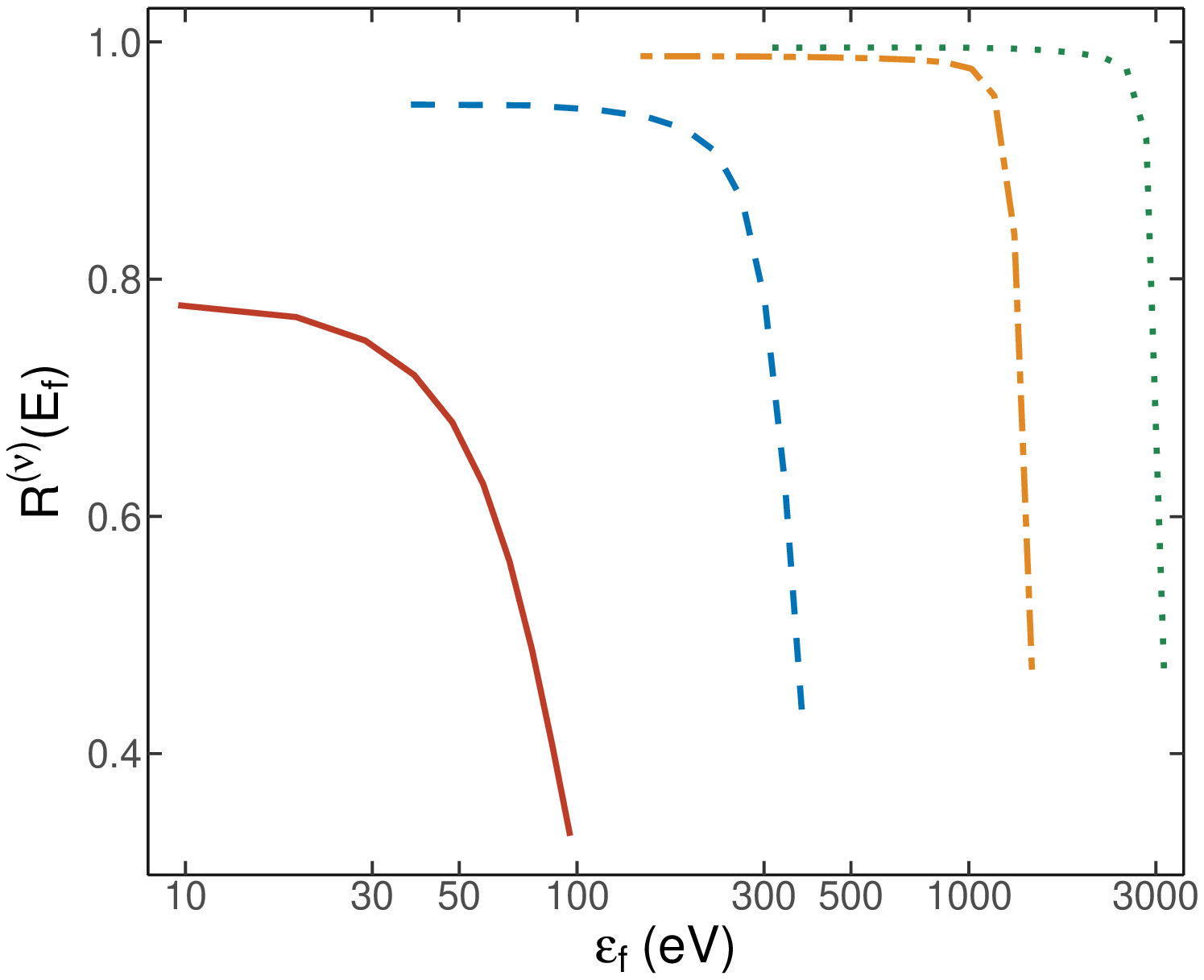}
\caption{\label{fig:Hydrogen}Energy spectra ratios  $R^{(\nu)}(E_{f})$ (Eqn (\ref{W100a})),  
as a function of electron kinetic energy $\epsilon_{f}$, 
of ionization electrons resulting from scattering of neutrinos by ground state hydrogen.
Results are shown for scattering of 5 keV (solid line), 10 keV (dashed line), 20 keV (dash-dotted line), 
and 30 keV (dotted line) incident neutrino energies. }
\end{figure}

\begin{figure}[ht]
\includegraphics[width=0.95\columnwidth]{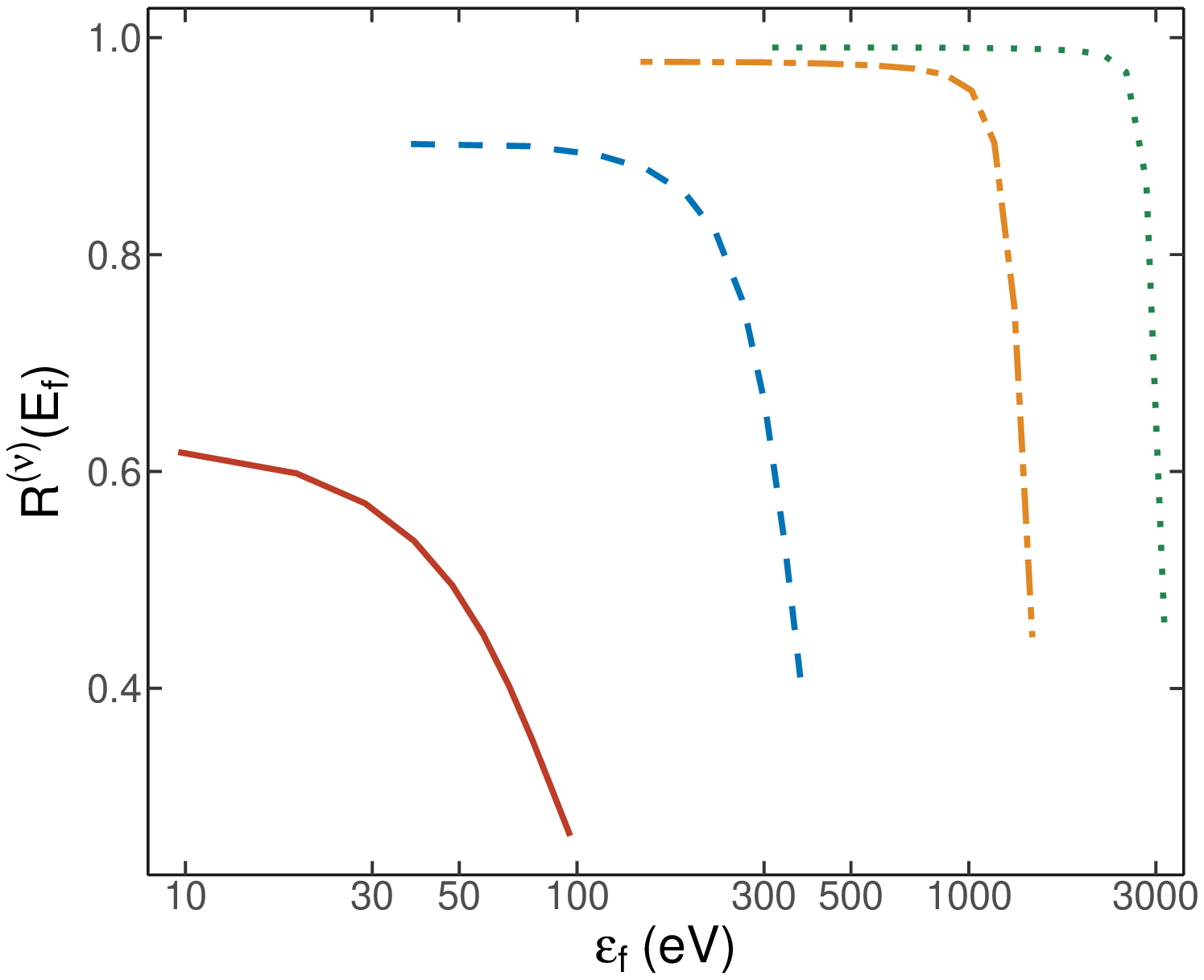}
\caption{\label{fig:Helium}Energy spectra ratios $R^{(\nu)}(E_{f})$ (Eqn (\ref{W100a})),   
as a function of electron kinetic energy $\epsilon_{f}$,
of ionization electrons resulting from scattering of neutrinos by ground state helium.
Results are shown for scattering of 5 keV (solid line), 10 keV (dashed line), 20 keV (dash-dotted line), 
and 30 keV (dotted line) incident neutrino energies. }
\end{figure}

\begin{figure}[ht]
\includegraphics[width=0.95\columnwidth]{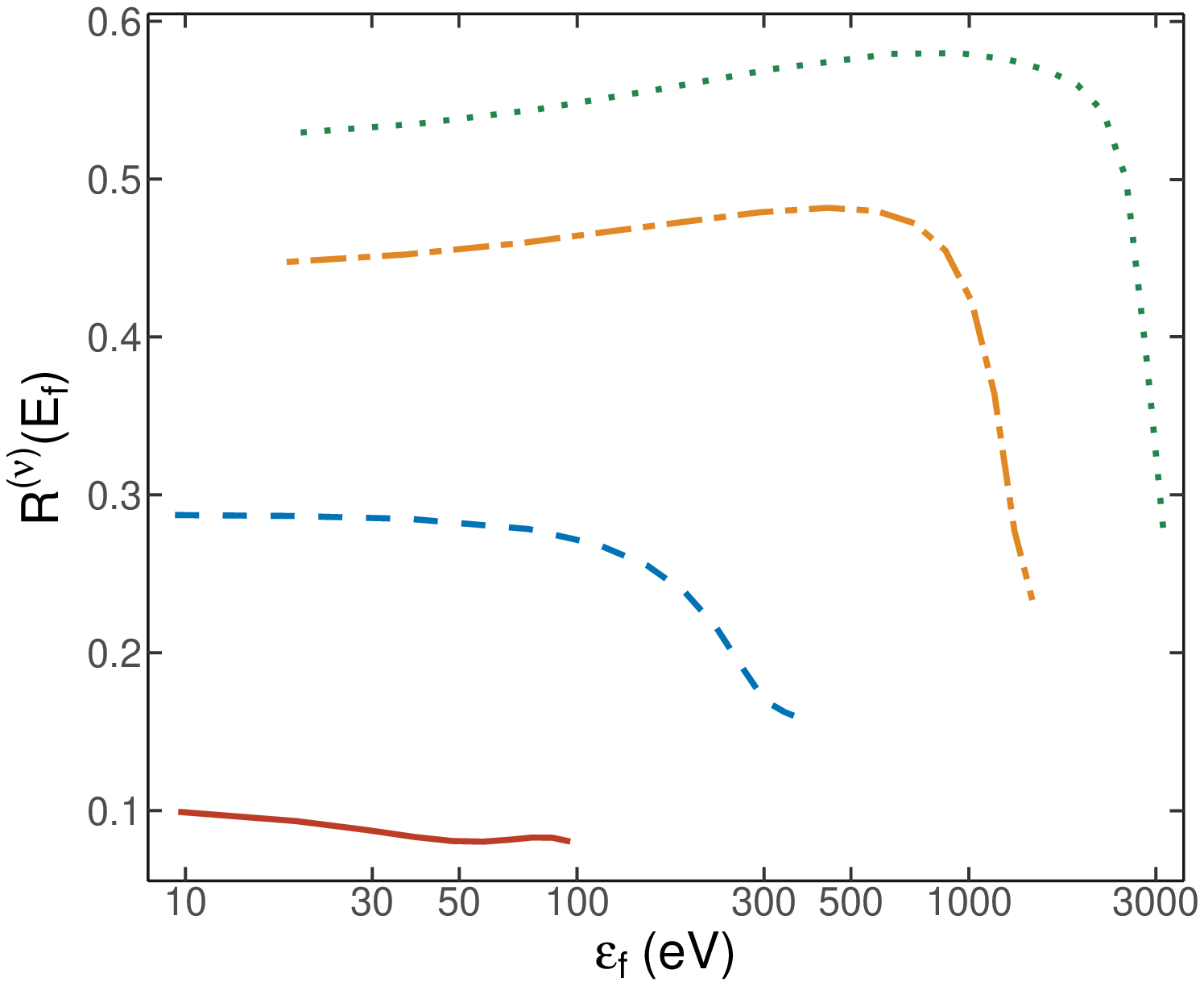}
\caption{\label{fig:Neon}Energy spectra ratios $R^{(\nu)}(E_{f})$ (Eqn (\ref{W100a})),  
as a function of electron kinetic energy $\epsilon_{f}$,
of ionization electrons resulting from scattering of neutrinos by ground state neon.
Results are shown for scattering of 5 keV (solid line), 10 keV (dashed line), 20 keV (dash-dotted line), 
and 30 keV (dotted line) incident neutrino energies. }
\end{figure}

The energy spectra of electrons resulting from the scattering of neutrinos by free electrons 
are shown in Figure  \ref{fig:free}, 
and the energy spectra ratios for scattering of neutrinos by H, He, and Ne are shown in 
Figs \ref{fig:Hydrogen}, \ref{fig:Helium}, and \ref{fig:Neon} respectively. 
Plots for scattering of antineutrinos differ only very slightly from those for
scattering by neutrinos and are not shown. The energy spectra ratios for scattering of 
10, 20, and 30  keV neutrinos by H, He, and Ne become constant at low kinetic energies and 
can safely be extrapolated to lower kinetic energies by assuming the ratios are constant.

The calculations involve a sum over $\kappa_{f}$ with convergence decreasing with increasing 
$\epsilon_{f}$.  The choice $|\kappa_{f}| \le 20$ for $E_{\nu_{i}}=5 $  keV,  
$|\kappa_{f}| \le 30$ for $E_{\nu_{i}}=10 $ keV, and $|\kappa_{f}| \le 50$ for
 $E_{\nu_{i}}=20 $ keV, gave convergence of much better than $1 \times 10^{-4}$ for each energy 
spectrum. For the larger electron kinetic energies at $E_{\nu_{i}}=30 $ keV, the imposed practical limit 
$|\kappa_{f}| \le 50$ gave convergence of better than $1 \times 10^{-4}$ for 
$\epsilon_{f} \alt 1.5 $ keV in the spectrum but, for higher energies, the convergence decreased 
to $1 \times 10^{-3}$ at $\epsilon_{f}^{\mathrm{max}}$, so the
numbers shown for the high energy end of the spectra are slight underestimates. For Ne, this 
decrease in convergence to below $1 \times 10^{-4}$ only occurred for the L$_{\mathrm{III}}$ subshell.

It is evident in the calculated spectra that binding effects increase strongly with atomic number, 
are greatest for low $E_{\nu_{i}}$ and, for each $E_{\nu_{i}}$, most significant at the 
high electron energy end of the spectrum. 
As expected, the binding effects are less for a free final electron than for a Coulombic final electron.
%With increasing $E_{\nu_{i}}$, the binding effects are pushed to a smaller and smaller region
%at the high energy end of the spectrum. 
For Ne, the binding effects were strongest for the K shell. The K shell results also showed the greatest 
enhancement from the use of Coulombic final electron states.
To a lesser extent, this was also the case for the L$_{\mathrm{I}}$ subshell at the lower neutrino energies..
 
The sharp decrease in the spectra at the high energy end is a consequence of the very small range of the
$q^2$ integration in (\ref{W70a}) for this region. Since $q_{\mathrm{max}} = E_{\nu_{i}} + E_{\nu_{f}}$ and 
$q_{\mathrm{min}}= E_{\nu_{i}}-E_{\nu_{f}}$, the range $2E_{\nu_{f}}$ is 
minimized at $E_{f}^{\mathrm{max}}$. Also, for this region, $q \approx E_{\nu_{i}}$, so that the 
high energy tail will increase as $E_{\nu_{i}}$ increases.

The shape of the Coulombic and free final electron energy spectra ratios $R(E_{f})$
differ slightly. The Coulombic final electron spectra ratios are approximately constant 
over the region just above $\epsilon_{f}^{\mathrm{min}} = 0.1 \epsilon_{f}^{\mathrm{max}} $
before decreasing, in most cases, monotonically, whereas
the free final electron spectra ratios increase initially with 
$\epsilon_{f}$ to a small peak before decreasing monotonically.

Existing calculations~\cite{Gounaris2002,Gounaris2004} model the scattering by a 
bound electron as scattering from a free electron with effective
mass $\tilde{m}$. The cross section obtained from (\ref{W03})  is
\begin{eqnarray}
\label{W104}
d \sigma^{(\mathrm{Atom})} & = & \frac{1}{4E_{\nu_{i}}E_{i}} 
\int \frac{d^{3}\mathbf{p}_{e_{i}}}{(2 \pi )^{3}}  |\Psi_{n_{i}l_{i}m_{i}}(\mathbf{p}_{e_{i}})|^{2}
\nonumber  \\ 
&& \times |F(\nu_{e}e^{-} \rightarrow \nu_{e}e^{-})|^{2} \,d\Phi^{(2)}(p_{e_{f}},p_{\nu_{f}}),
\end{eqnarray}
where the two-body phase space is
\begin{equation}
\label{W105}
d \Phi^{(2)}(p_{e_{f}},p_{\nu_{f}}) = \frac{1}{8 \pi} \frac{1}{s-\tilde{m}^{2}} dt.
\end{equation}
The ejected electron energy spectrum is then
\begin{eqnarray}
\label{W106}
\frac{d \sigma^{(\mathrm{Atom})}}{d E_{f}} & = & \frac{1}{32 \pi E_{\nu_{i}}E_{i}} 
\int \frac{d^{3}\mathbf{p}_{e_{i}}}{(2 \pi )^{3}}  |\Psi_{n_{i}l_{i}m_{i}}(\mathbf{p}_{e_{i}})|^{2}
\nonumber  \\ 
&& \times |F(\nu_{e}e^{-} \rightarrow \nu_{e}e^{-})|^{2} \,\frac{1}{s-\tilde{m}^{2}} \frac{dt}{d E_{f}}.
\end{eqnarray}
In the rest frame of the atom, with the incoming neutrino along $Oz$ and the outgoing electron 
lying in the $Oxz$ plane,
\begin{equation}
\label{W107}
s = \tilde{m}^{2}+2E_{\nu_{i}}(E_{i}-p_{e_{i}} \cos \theta_{e_{i}})
\end{equation}
and 
\begin{equation}
\label{W108}
t = 2 E_{\nu_{i}}(E_{f}-p_{e_{f}} \cos \theta_{e_{f}} -E_{i}-p_{e_{i}} \cos \theta_{e_{i}}).
\end{equation}
The collision kinematics restrict the range of $E_{f}$ to 
$E_{f}^{(2)} \le E_{f} \le E_{f}^{(1)}$ where~\cite{Gounaris2004} the limits $E_{f}^{(1,2)}$
depend on $\mathbf{p}_{e_{i}}$.

The energy spectra ratios calculated by~\cite{Gounaris2004} have a similar shape to the free final electron spectra 
ratios calculated here in that the ratios increase initially to a small peak before decreasing monotonically.
The present free final electron ratios, however, differ significantly in magnitude to those of~\cite{Gounaris2004}
and have a much smoother energy dependence.

The integrated cross sections
\begin{equation}
\label{W109a}
\sigma^{(\nu)} = \int_{E_{l}}^{E_{u}} dE_{f}  \, \frac{d\sigma^{(\nu)} }{dE_{f}},
\end{equation}
where $E_{l}=m_{e}$ and $E_{u}= m_{e}+\epsilon_{f}^{\mathrm{max}}$, can be estimated from the 
calculated energy spectra. As these spectra were only calculated for 
$E_{f} \ge m_{e}+ 0.1 \epsilon_{f}^{\mathrm{max}}$, we assume the spectra at $E_{f}=m_{e}$
are the same as at $m_{e}+ 0.1 \epsilon_{f}^{\mathrm{max}}$. This assumes the lower energy
part of the spectrum is flat.
These integrated  cross sections are
given in the tables, expressed as ratios to the integrated cross sections 
$Z\sigma^{(\nu)}_{(\mathrm{Free})}$ (Eqn (\ref{W102a})) for $Z$ free electrons.

\section{Summary and conclusions}

The energy spectra $d \sigma/dE_{f}$ of the ionization electrons produced in the scattering
of electron neutrinos and antineutrinos with energies 5, 10, 20 and 30 keV by atomic electrons 
have been calculated for scattering by the ground state systems H($1s$), He($1s^{2}$), 
and Ne($1s^{2}2s^{2}2p^{6}$) where, for He and Ne, 
the electrons are considered as independent scattering centers. Results are also obtained for the 
integrated cross sections.

The present calculations maintain the full collision dynamics by formulating the scattering  
in configuration space using the Bound Interaction 
Picture, rather than the usual formulation in the Interaction Picture in momentum space as 
appropriate to scattering by free electrons. The energy spectra are expressed as an integral
over the momentum transfer $q$ from the neutrino or antineutrino to the atomic system. The integrand is the
contraction of the second rank neutrino tensor $L^{(\nu)}$ (eqn (\ref{W44})) or antineutrino tensor 
$L^{(\bar{\nu})}$ (eqn (\ref{W44a})), and the second rank electron tensor $L^{(e)}$ (eqn (\ref{W68})).
This electron tensor involves radial integrals over Dirac central field radial eigenfunctions 
for the initial bound electron and final continuum electron, together with a spherical Bessel function
arising from the momentum transfer. Screened point-Coulomb radial eigenfunctions have been used,
with the continuum state eigenfunctions calculated by direct integration of the Dirac equations.

The calculated energy spectra have been expressed as ratios to the energy spectra for scattering by free electrons.
Binding effects increase strongly with atomic number, are largest for low $E_{\nu_{i}}$ and, 
for each $E_{\nu_{i}}$, greatest at the high electron energy end of the spectrum. The most
extreme effects of binding are for $E_{\nu_{i}} = 5$ keV scattering by Ne where the ratios are less
than $0.1$. The energy spectra have been calculated for both a Coulombic final electron state and a free final electron state.
The results indicate that the binding effects from the continuum state of the final electron are 
significant and can be comparable to those arising from the bound initial electron state. This 
especially occurs at the high energy end of the spectra for scattering of 5 and 10 keV neutrinos and
antineutrinos. As the continuum radial eigenfunctions increase as $r^{\gamma_{f}}$ until the point 
$r_{s}\approx |\kappa_{f}|/p_{f}$ where they become oscillatory (see Appendix C), the continuum
state for high $E_{f}$ and $p_{f}$ contributes strongly at small distances where the Coulomb 
interaction is significant. All existing calculations assume a free final electron and therefore under 
estimate the total binding effects.

The neutrino and antineutrino energy spectra are very similar, with the small difference of 
$\alt 1\% $ arising from the $(1,2)$ and $(2,1)$ elements of the lepton tensor $L^{(\nu)}{}^{\beta ,\alpha}$.
In all cases the free electron and bound electron energy spectra for neutrino scattering are higher 
than those for antineutrinos, although the ratios at the high energy end of the spectra do not reflect this.

The results for the Ne energy spectra show that binding effects are still very significant at $E_{\nu_{i}} = 30$ keV,
the integrated spectra ratios being $\alt 0.6$. This suggests
the calculations should be extended to higher neutrino energies for this element.  However,
this would require a substantial increase in the maximum value of $|\kappa_{f}|$ used as the $\kappa_{f}$
convergence is very slow at higher energies, which is not practical. As the results form a monotonic 
sequence for increasing values of $\kappa_{f}$, a convergence
acceleration technique such as the $\theta$-algorithm~\cite{Wimp1981} may be beneficial.

The formalism and techniques developed in the present calculations have assumed the atomic electrons
for $Z \ne 1$ are in a closed shell or subshell (see eqn (\ref{W63})). They can be readily applied to 
other closed shell or subshell atomic systems provided these systems can be represented by central field eigenfunctions. 
Only the calculation of the radial integrals 
would require modification to deal with non-Coulombic self-consistent relativistic radial eigenfunctions.

\appendix

\section{Scattering matrix}

The second order $S$ operator for the interaction (\ref{W1}) between an electron-neutrino 
and an electron is
\begin{equation}
\label{W5}
S^{(2)}=-\frac{1}{2} \int d^{4}x_{1} \int d^{4}x_{2} \,
T \{\mathcal{L}^{\nu_{e}e}_{I} (x_{1})\mathcal{L}^{\nu_{e}e}_{I}(x_{2}) \},
\end{equation}
where $T$ is the time ordering operator. Of relevance to $\nu_{e} -e$ scattering are the terms
\begin{widetext}
\begin{equation}
\label{W6}
S^{(2)}_{\nu_{e}e}  =  -\frac{1}{2} \int d^{4}x_{1} \int d^{4}x_{2} 
T \; \{\mathcal{L}^{\nu_{e}We}_{I}(x_{1}) \mathcal{L}^{\nu_{e}We}_{I}(x_{2})  
+ \mathcal{L}^{\nu_{e}Z\nu_{e}}_{I} (x_{1}) \mathcal{L}^{eZe}_{I}(x_{2})
+ \mathcal{L}^{eZe}_{I}(x_{1}) \mathcal{L}^{\nu_{e}Z\nu_{e}}_{I}(x_{2})\}. 
\end{equation}
Using Wick's theorem, the three time-ordered terms become
\begin{eqnarray}
T \{\mathcal{L}^{\nu_{e}We}_{I}(x_{1}) \mathcal{L}^{\nu_{e}We}_{I}(x_{2})\} &= &
\left(\frac{-g}{2\sqrt{2}} \right)^{2} 
\{ N[\bar{\nu_{e}}(x_{1})\gamma^{\alpha}(1-\gamma_{5}) e(x_{1})
\bcontraction{}{W^{(+)}_{\alpha} (x_{1})}{}{ W^{(-)}_{\beta}(x_{2})}
W^{(+)}_{\alpha} (x_{1}) W^{(-)}_{\beta}(x_{2})
%\underbrace{W^{(+)}_{\alpha} (x_{1}) W^{(-)}_{\beta}(x_{2})}
\bar{e}(x_{2}) \gamma^{\beta}(1-\gamma_{5})\nu_{e}(x_{2})] \nonumber  \\
&& + (x_{1} \leftrightarrow x_{2}) \}
\label{W7a} \\
T \{\mathcal{L}^{\nu_{e}Z\nu_{e}}_{I} (x_{1}) \mathcal{L}^{eZe}_{I}(x_{2})\} & = &
\left( \frac{-g}{4 \cos \theta_{W}}\right)^{2}
N[\bar{\nu_{e}}(x_{1}) \gamma^{\alpha}(1-\gamma_{5})\nu_{e}(x_{1})
\bcontraction{}{Z_{\alpha} (x_{1})}{}{Z_{\beta}(x_{2})}
Z_{\alpha} (x_{1}) Z_{\beta}(x_{2})
%\underbrace{Z_{\alpha} (x_{1}) Z_{\beta}(x_{2})}
\bar{e}(x_{2}) \gamma^{\beta}(v_{e}+a_{e}\gamma_{5}) e (x_{2}) ]  \label{W7b} \\
T \{\mathcal{L}^{eZe}_{I}(x_{2}) \mathcal{L}^{\nu_{e}Z\nu_{e}}_{I} (x_{1})\} & = & 
T \{\mathcal{L}^{\nu_{e}Z\nu_{e}}_{I} (x_{1}) \mathcal{L}^{eZe}_{I}(x_{2})\} 
(x_{1} \leftrightarrow x_{2})  \label{W7c}.
\end{eqnarray}
\end{widetext}
Here the contracted operators 
$
\bcontraction{}{W^{(+)}_{\alpha} (x_{1})}{}{ W^{(-)}_{\beta}(x_{2})}
W^{(+)}_{\alpha} (x_{1}) W^{(-)}_{\beta}(x_{2})
$ 
and
$
\bcontraction{}{Z_{\alpha} (x_{1})}{}{Z_{\beta}(x_{2})}
Z_{\alpha} (x_{1}) Z_{\beta}(x_{2})
$
are the $W$ and $Z$ gauge boson propagators
respectively, which we denote by $iD^{A}_{\alpha\beta}(x_{1},x_{2})$ with $A=W,Z$. As $x_{1}$ 
and $x_{2}$ are dummy integration variables, the interchanges in (\ref{W7a}) and (\ref{W7c})
produce terms identical to the original terms.

In order to identify the terms in (\ref{W7a}) and (\ref{W7b}) specific to the scattering of electron neutrinos 
by electrons, the neutrino and electron field operators are expanded in terms of appropriate 
basis sets of states, with the coefficients identified as the corresponding particle and 
antiparticle creation and annihilation operators. For the neutrino field we have the usual 
expansion in terms of neutrino $u^{(s)}(\mathbf{p})$ and antineutrino $v^{(s)}(\mathbf{p})$
plane wave spinors:
\begin{equation}
\label{W8}
\nu_{e}(x)=\nu_{e}^{(+)}(x) + \nu_{e}^{(-)}(x),
\end{equation}
where
\begin{equation}
\label{W9}
\nu_{e}^{(+)}(x)=\frac{1}{(2 \pi )^{3}} \sum_{s=\pm 1/2} \int d^{3}p \frac{1}{2E_{\nu}}
u^{(s)}(\mathbf{p}) b_{s}(\mathbf{p}) e^{-ip\cdot x} 
\end{equation}
and
\begin{equation}
\label{W10}
\nu_{e}^{(-)}(x)=\frac{1}{(2 \pi )^{3}} \sum_{s=\pm 1/2} \int d^{3}p \frac{1}{2E_{\nu}}
v^{(s)}(\mathbf{p}) d_{s}^{\dagger}(\mathbf{p}) e^{ip\cdot x} .
\end{equation}
The expansions for the conjugate field $\bar{\nu_{e}}\equiv \nu_{e}^{\dagger}\gamma^{0}$ are
\begin{equation}
\label{W11}
\bar{\nu_{e}}(x)=\bar{\nu_{e}}^{(+)}(x) + \bar{\nu_{e}}^{(-)}(x),
\end{equation}
where
\begin{equation}
\label{W12}
\bar{\nu_{e}}^{(+)}(x)=\frac{1}{(2 \pi )^{3}} \sum_{s=\pm 1/2} \int d^{3}p \frac{1}{2E_{\nu}}
\bar{v}^{(s)}(\mathbf{p}) d_{s}(\mathbf{p}) e^{-ip\cdot x} .
\end{equation}
and
\begin{equation}
\label{W13}
\bar{\nu_{e}}^{(-)}(x)=\frac{1}{(2 \pi )^{3}} \sum_{s=\pm 1/2} \int d^{3}p \frac{1}{2E_{\nu}}
\bar{u}^{(s)}(\mathbf{p}) b_{s}^{\dagger}(\mathbf{p}) e^{ip\cdot x}. 
\end{equation}
In the above, $b_{s}(\mathbf{p})$ ($d_{s}(\mathbf{p})$) are the neutrino (antineutrino) 
one-particle annihilation operators, and $b_{s}^{\dagger}(\mathbf{p})$ 
($d_{s}^{\dagger}(\mathbf{p})$) the neutrino (antineutrino) one-particle creation operators,
for particles with momentum $\mathbf{p}$ and helicity $s$. Choosing the normalization
\begin{eqnarray}
\label{W13a}
\bar{u}^{(s^{\prime})}(\mathbf{p}) \gamma^{0}u^{(s)}(\mathbf{p}) & = & 2p^{0} \delta_{s^{\prime},s},
\nonumber  \\
\bar{v}^{(s^{\prime})}(\mathbf{p}) \gamma^{0}v^{(s)}(\mathbf{p}) & = & 2p^{0} \delta_{s^{\prime},s},
\end{eqnarray}
then the annihilation and creation operators satisfy
\begin{eqnarray}
\label{W13b}
\{ b_{s}(\mathbf{p}), b_{s^{\prime}}^{\dagger}(\mathbf{p}^{\prime}\} & =&
(2 \pi )^{3} 2 p^{0} \delta_{s^{\prime},s} \delta (\mathbf{p}-\mathbf{p}^{\prime}),
\nonumber  \\
\{ d_{s}(\mathbf{p}), d_{s^{\prime}}^{\dagger}(\mathbf{p}^{\prime}\} & =&
(2 \pi )^{3} 2 p^{0} \delta_{s^{\prime},s} \delta (\mathbf{p}-\mathbf{p}^{\prime}).
\end{eqnarray}

For the electron field, however, the decomposition is in terms of solutions $\phi(x)$ of (\ref{W0}) 
for an electron in the external field $A^{(\mathrm{ext})}$. Assuming the positive and negative 
energy solutions $\phi_{n}^{(\pm)}(x)$ form two distinct sets, each separated from the zero 
energy by a finite interval, we can make the expansion ($E_{n}>0$)
\begin{equation}
\label{W14}
e(x) = e^{(+)}(x) + e^{(-)}(x),
\end{equation}
where
\begin{equation}
\label{W15}
e^{(+)}(x) = \sum_{n} a_{n}  \phi_{n}^{(-)}(\mathbf{x})  e^{-i E_{n}x^{0}},
\end{equation}
and
\begin{equation}
\label{W16}
e^{(-)}(x) = \sum_{n} c_{n}  \phi_{n}^{(+)}(\mathbf{x})  e^{i E_{n}x^{0}}.
\end{equation}
Similarly, for the conjugate field,
\begin{equation}
\label{W17}
\bar{e}(x) = \bar{e}^{(+)}(x) + \bar{e}^{(-)}(x),
\end{equation}
where
\begin{equation}
\label{W19}
\bar{e}^{(+)}(x) = \sum_{n} c^{\dagger}_{n}  \bar{\phi}_{n}^{(+)}(\mathbf{x})  e^{-i E_{n}x^{0}},
\end{equation}
and
\begin{equation}
\label{W18}
\bar{e}^{(-)}(x) = \sum_{n} a^{\dagger}_{n}  \bar{\phi}_{n}^{(-)}(\mathbf{x})  e^{i E_{n}x^{0}}.
\end{equation}
Here, $n$ represents the set of quantum numbers, including $E_{n}$, specifying the states
in the external field, $a_{n}$ ($c_{n}$) are annihilation operators for electrons with positive (negative) energies,
and $a^{\dagger}_{n}$ ($c^{\dagger}_{n}$) are the corresponding creation operators. If the external-field
solutions are normalized according to 
\begin{equation}
\label{W18a}
\langle \phi^{(\pm)}_{n^{\prime}} | \phi^{(\pm)}_{n} \rangle = \delta_{n^{\prime},n},
\end{equation}
then these operators satisfy
\begin{equation}
\label{W18b}
\{a_{n}, a_{n^{\prime}}^{\dagger}\} =\{c_{n}, c_{n^{\prime}}^{\dagger}\} = \delta_{n^{\prime},n}.
\end{equation} 

The relevant terms in (\ref{W7a}) are
\begin{eqnarray}
\label{W19b}
T^{W}(x_{1},x_{2}) & = & N [\bar{\nu_{e}}^{(\mp)}(x_{1})V^{\alpha}e^{(+)}(x_{1}) 
iD^{W}_{\alpha \beta }(x_{1},x_{2})  \nonumber  \\
&& \times  \bar{e}^{(-)}(x_{2}) V^{\beta} \nu_{e}^{(\pm)} (x_{2})] ,
\end{eqnarray}
where we have defined the vertex factor
\begin{equation}
\label{W19a}
V^{\alpha} \equiv \frac{-g}{2 \sqrt{2}} \gamma^{\alpha} (1-\gamma_{5}).
\end{equation}
From (\ref{W7b}), the terms are
\begin{eqnarray}
\label{W20}
T^{Z}(x_{1},x_{2}) & = & N [\bar{\nu_{e}}^{(\mp)}(x_{1})V^{\alpha}_{\nu} 
\nu _{e}^{(\pm)}(x_{1}) iD^{Z}_{\alpha \beta} (x_{1},x_{2})  \nonumber  \\
&& \times \bar{e}^{(-)}(x_{2})V^{\beta}_{e} e^{(+)}(x_{2}) ],
\end{eqnarray}
where the vertex factors are
\begin{eqnarray}
V^{\alpha}_{\nu} &\equiv &  \frac{-g}{4 \cos \theta_{W}} \gamma^{\alpha}(1-\gamma_{5}), 
\label{W20a} \\
V^{\beta}_{e} & \equiv & \frac{-g}{4 \cos \theta_{W}} \gamma^{\beta}(v_{e}+a_{e}\gamma_{5}).
\label{W20b}
\end{eqnarray}
The upper (lower) signs on the neutrino field operators in (\ref{W19b}) and (\ref{W20}) refer 
to $\nu_{e}$ ($\bar{\nu_{e}}$) scattering.

Introducing the Fourier transform of the gauge boson propagator
\begin{equation}
\label{W21}
D^{A}_{\alpha \beta} (x_{1},x_{2}) =\frac{1}{(2 \pi)^{4}} \int d^{4}k \,
D^{A}_{\alpha \beta} (k) e^{ik\cdot(x_{1}-x_{2})},
\end{equation}
where
\begin{equation}
\label{W22}
iD^{A}_{\alpha \beta} (k) = - \frac{i}{k^{2}-M_{A}^{2}-i \epsilon} 
\left[g_{\alpha \beta} - \frac{(1-\xi)k_{\alpha}k_{\beta}}{k^{2}-\xi M_{A}^{2}} \right],
\end{equation}
and substituting the expansions (\ref{W9}), (\ref{W13}), (\ref{W15}) and (\ref{W18})
for the field operators in (\ref{W19}) allows the integrations over $d x_{1}^{0},d x_{2}^{0}$
and $d k^{0}$ to be performed, yielding the energy conservation condition 
$\delta(E_{n}^{\prime}+E_{\nu}-E_{n}-E_{\nu}^{\prime})$ and the $S$-operator 
for $W$ mediated $\nu_{e}$ scattering
\begin{widetext}
\begin{eqnarray}
\label{W23}
S^{(W,\nu)} & = & -\frac{1}{(2 \pi )^{8}} \sum_{n^{\prime},n} \sum_{s^{\prime},s}
\;\int d^{3} p^{\prime} \int d^{3} p \int d^{3} k \frac{1}{4 E_{\nu}^{\prime}E_{\nu}} 
\delta(E_{n}^{\prime}+E_{\nu}-E_{n}-E_{\nu}^{\prime}) 
\int d^{3}x_{1} \int d^{3} x_{2} e^{-i(\mathbf{p}^{\prime}+\mathbf{k})\cdot \mathbf{x}_{1}}
e^{i(\mathbf{p}+\mathbf{k})\cdot \mathbf{x}_{2}}  \nonumber  \\
&& \times \bar{u}^{(s^{\prime})}(\mathbf{p}^{\prime})V^{\alpha} \phi^{(+)}_{n}(\mathbf{x}_{1})
i D^{W}_{\alpha \beta}(k^{0},\mathbf{k}) 
\bar{\phi}^{(+)}_{n^{\prime}} (\mathbf{x}_{2}) V^{\beta} u^{(s)}(\mathbf{p})
N[b^{\dagger}_{s^{\prime}}(\mathbf{p}^{\prime})a_{n^{\prime}} a^{\dagger}_{n} b_{s}(\mathbf{p})],
\end{eqnarray}
where $k^{0}=E_{n}-E_{\nu}$. The same substitutions into (\ref{W20}) allows the 
integrations over $d^{4}x_{1}, d^{4}k$ and $d x^{0}_{2}$
to be performed, resulting in the $S$-operator for $Z$-mediated $\nu_{e}$ scattering
\begin{eqnarray}
\label{W24}
S^{(Z,\nu)} & = & - \frac{1}{(2 \pi )^{5}} \sum_{n^{\prime},n} \sum_{s^{\prime},s}
\;\int d^{3} p^{\prime} \int d^{3} p  \frac{1}{4 E_{\nu}^{\prime}E_{\nu}} 
\delta(E_{n}^{\prime}+E_{\nu}^{\prime}-E_{n}-E_{\nu}) 
\int d^{3}x_{2} e^{i(\mathbf{p}-\mathbf{p}^{\prime})\cdot \mathbf{x}_{2}}
\nonumber  \\
&& \times  \bar{u}^{(s^{\prime})}(\mathbf{p}^{\prime})V^{\alpha}_{\nu} u^{(s)}(\mathbf{p}) 
i D^{Z}_{\alpha \beta}(p^{\prime}-p) 
\bar{\phi}^{(+)}_{n^{\prime}} (\mathbf{x}_{2}) V^{\alpha}_{e}\phi^{(+)}_{n}(\mathbf{x}_{2})
N[b^{\dagger}_{s^{\prime}}(\mathbf{p}^{\prime})b_{s}(\mathbf{p})a^{\dagger}_{n^{\prime}} a_{n}].
\end{eqnarray}
\end{widetext}
Forming the $S$-matrix between the initial state
\begin{equation}
\label{W25}
|i \rangle = b^{\dagger}_{s_{i}}(\mathbf{p}_{\nu_{i}}) a^{\dagger}_{n_{i}} |\mathrm{Vac} \rangle
\end{equation}
and the final state
\begin{equation}
\label{W26}
|f \rangle = b^{\dagger}_{s_{f}}(\mathbf{p}_{\nu_{f}}) a^{\dagger}_{n_{f}} |\mathrm{Vac} \rangle
\end{equation}
gives
\begin{widetext}
\begin{eqnarray}
\label{W27}
S^{(W,\nu)}_{fi} & = & -\frac{1}{(2 \pi )^{2}} \delta(E_{fi}^{(\nu)}) \int d^{3} k \int d^{3}x_{1} \int d^{3} x_{2} 
e^{-i(\mathbf{p}_{\nu_{f}}+\mathbf{k})\cdot \mathbf{x}_{1}}
e^{i(\mathbf{p}_{\nu_{i}}+\mathbf{k})\cdot \mathbf{x}_{2}} 
\nonumber  \\ 
&& \times \bar{u}^{(s_{f})}(\mathbf{p}_{\nu_{f}})V^{\alpha} \phi^{(+)}_{n_{i}}(\mathbf{x}_{1})
i D^{W}_{\alpha \beta}(k^{0},\mathbf{k}) 
\bar{\phi}^{(+)}_{n_{f}} (\mathbf{x}_{2}) V^{\beta} u^{(s_{i})}(\mathbf{p}_{\nu_{i}})
\end{eqnarray}
and
\begin{equation}
\label{W28}
S^{(Z,\nu)}_{fi}  =- 2 \pi \delta(E_{fi}^{(\nu)}) \int d^{3}x_{2} 
e^{i(\mathbf{p}_{\nu_{i}}-\mathbf{p}_{\nu_{f}})\cdot \mathbf{x}_{2}}
\bar{u}^{(s_{f})}(\mathbf{p}_{\nu_{f}})V^{\alpha}_{\nu} u^{(s_{i})}(\mathbf{p}_{\nu_{i}})
i D^{Z}_{\alpha \beta}(p_{i}-p_{f})
\bar{\phi}^{(+)}_{n_{f}} (\mathbf{x}_{2}) V^{\alpha}_{e}\phi^{(+)}_{n_{i}}(\mathbf{x}_{2}),
\end{equation}
\end{widetext}
where
\begin{equation}
\label{W29}
\delta(E_{fi}^{(\nu)}) \equiv \delta(E_{n_{f}}+E_{\nu_{f}}-E_{n_{i}}-E_{\nu_{i}}) .
\end{equation}

Since the scattering occurs at low momentum transfers $k^{2} \ll M_{A}^{2}$, the gauge boson
propagator, in the Feynman gauge $\xi =1$, simplifies to
\begin{equation}
\label{W30}
i D^{A}_{\alpha \beta }(k)=\frac{i}{M^{2}_{A}} g_{\alpha \beta},
\end{equation}
and the integrations over $d^{3} k$ and $d^{3} x_{1}$ in (\ref{W27}) can now be performed. Noting that
$M_{W}=M_{Z} \cos \theta_{W}$ and $g^{2}/(8 M_{W}^{2})=G_{\mathrm{F}}/\sqrt{2}$, the 
$S$-matrix elements become
\begin{eqnarray}
\label{W31}
S^{(W,\nu)}_{fi} & = & - 2 \pi i \frac{G_{\mathrm{F}}}{\sqrt{2}} \delta(E_{fi}^{(\nu)}) \int d^{3}x_{2}  
e^{i(\mathbf{p}_{\nu_{i}}-\mathbf{p}_{\nu_{f}})\cdot \mathbf{x}_{2}}  
\nonumber  \\
&& \times 
[\bar{u}^{(s_{f})}(\mathbf{p}_{\nu_{f}})\gamma^{\alpha}(1-\gamma_{5}) \phi^{(+)}_{n_{i}}(\mathbf{x}_{2})]
\nonumber  \\
&& \times [\bar{\phi}^{(+)}_{n_{f}} (\mathbf{x}_{2}) \gamma_{\alpha}(1-\gamma_{5}) u^{(s_{i})}(\mathbf{p}_{\nu_{i}})],
\end{eqnarray}
and
\begin{eqnarray}
\label{W32}
S^{(Z,\nu)}_{fi} & = & -\pi i \frac{G_{\mathrm{F}}}{\sqrt{2}}\delta(E_{fi}^{(\nu)}) \int d^{3}x_{2}  
e^{i(\mathbf{p}_{\nu_{i}}-\mathbf{p}_{\nu_{f}})\cdot \mathbf{x}_{2}}  
\nonumber  \\
&& \times [\bar{u}^{(s_{f})}(\mathbf{p}_{\nu_{f}})\gamma^{\alpha}(1-\gamma_{5}) u^{(s_{i})}(\mathbf{p}_{\nu_{i}})]
\nonumber  \\
&& \times
[\bar{\phi}^{(+)}_{n_{f}} (\mathbf{x}_{2}) \gamma_{\alpha}(v_{e}+a_{e}\gamma_{5})\phi^{(+)}_{n_{i}}(\mathbf{x}_{2})].
\nonumber  \\
\end{eqnarray}
Using the Fierz transformation
\begin{eqnarray}
\label{W33}
[\bar{u}^{(s_{f})}(\mathbf{p}_{\nu_{f}})\gamma^{\alpha}(1-\gamma_{5}) \phi^{(+)}_{n_{i}}(\mathbf{x}_{2})] 
\nonumber  \\
\times [\bar{\phi}^{(+)}_{n_{f}} (\mathbf{x}_{2}) \gamma_{\alpha}(v_{e}+a_{e}\gamma_{5}) u^{(s_{i})}(\mathbf{p}_{\nu_{i}})]
\nonumber  \\
= [\bar{u}^{(s_{f})}(\mathbf{p}_{\nu_{f}})\gamma^{\alpha}(1-\gamma_{5}) u^{(s_{i})}(\mathbf{p}_{\nu_{i}})]
\nonumber  \\
\times [\bar{\phi}^{(+)}_{n_{f}} (\mathbf{x}_{1}) \gamma_{\alpha}(v_{e}+a_{e}\gamma_{5})\phi^{(+)}_{n_{i}}(\mathbf{x}_{1})]
\nonumber  \\
\end{eqnarray}
we can combine the $S$-matrices for $W$- and $Z$ - mediated scattering
to give the result (\ref{W34}). 

\section{Explicit expressions for electron scattering tensors}

Explicit expressions for the electron scattering tensors $L^{\beta \alpha}_{v_{e}v_{e}}, L^{\beta \alpha}_{a_{e}a_{e}}, 
L^{\beta \alpha}_{v_{e}a_{e}}$ and $L^{\beta \alpha}_{a_{e}v_{e}}$ appearing in (\ref{W68}) are given here. 
For the case $(\alpha, \beta)=(0,0)$, we have
\begin{widetext}
\begin{eqnarray}
L^{0,0}_{v_{e}v_{e}} & = & I^{gg*}_{\bar{l}}I^{gg}_{l}\tilde{A}^{0,0}_{\bar{l}l}(l_{f},l_{i},l_{f},l_{i}) +
I^{gg*}_{\bar{l}}I^{ff}_{l}\tilde{A}^{0,0}_{\bar{l}l}(l_{f},l_{i},l_{f}^{\prime},l_{i}^{\prime}) +
I^{ff*}_{\bar{l}}I^{gg}_{l}\tilde{A}^{0,0}_{\bar{l}l}(l_{f}^{\prime},l_{i}^{\prime},l_{f},l_{i}) +
I^{ff*}_{\bar{l}}I^{ff}_{l}\tilde{A}^{0,0}_{\bar{l}l}(l_{f}^{\prime},l_{i}^{\prime},l_{f}^{\prime},l_{i}^{\prime}),
\nonumber  \\
%\\
L^{0,0}_{a_{e}a_{e}} & = & I^{gf*}_{\bar{l}}I^{gf}_{l}\tilde{A}^{0,0}_{\bar{l}l}(l_{f},l_{i}^{\prime},l_{f},l_{i}^{\prime}) -
I^{gf*}_{\bar{l}}I^{fg}_{l}\tilde{A}^{0,0}_{\bar{l}l}(l_{f},l_{i}^{\prime},l_{f}^{\prime},l_{i}) -
I^{fg*}_{\bar{l}}I^{gf}_{l}\tilde{A}^{0,0}_{\bar{l}l}(l_{f}^{\prime},l_{i},l_{f},l_{i}^{\prime}) +
I^{fg*}_{\bar{l}}I^{fg}_{l}\tilde{A}^{0,0}_{\bar{l}l}(l_{f}^{\prime},l_{i},l_{f}^{\prime},l_{i}),
\nonumber  \\
%\\
L^{0,0}_{v_{e}a_{e}} & = & i [I^{gg*}_{\bar{l}}I^{gf}_{l}\tilde{A}^{0,0}_{\bar{l}l}(l_{f},l_{i},l_{f},l_{i}^{\prime}) -
I^{gg*}_{\bar{l}}I^{fg}_{l}\tilde{A}^{0,0}_{\bar{l}l}(l_{f},l_{i},l_{f}^{\prime},l_{i}) +
I^{ff*}_{\bar{l}}I^{gf}_{l}\tilde{A}^{0,0}_{\bar{l}l}(l_{f}^{\prime},l_{i}^{\prime},l_{f},l_{i}^{\prime}) -
I^{ff*}_{\bar{l}}I^{fg}_{l}\tilde{A}^{0,0}_{\bar{l}l}(l_{f}^{\prime},l_{i}^{\prime},l_{f}^{\prime},l_{i})],
\nonumber  \\
%\\
L^{0,0}_{a_{e}v_{e}} & = & -i [I^{gf*}_{\bar{l}}I^{gg}_{l}\tilde{A}^{0,0}_{\bar{l}l}(l_{f},l_{i}^{\prime},l_{f},l_{i}) +
I^{gf*}_{\bar{l}}I^{ff}_{l}\tilde{A}^{0,0}_{\bar{l}l}(l_{f},l_{i}^{\prime},l_{f}^{\prime},l_{i}^{\prime}) -
I^{fg*}_{\bar{l}}I^{gg}_{l}\tilde{A}^{0,0}_{\bar{l}l}(l_{f}^{\prime},l_{i},l_{f},l_{i}) -
I^{fg*}_{\bar{l}}I^{ff}_{l}\tilde{A}^{0,0}_{\bar{l}l}(l_{f}^{\prime},l_{i},l_{f}^{\prime},l_{i}^{\prime})]. 
\nonumber \\
\end{eqnarray}
\end{widetext}

The other cases can be obtained through the substitutions
\begin{eqnarray}
L^{k^{\prime},k}_{v_{e}v_{e}} & = & L^{0,0}_{a_{e}a_{e}}[\tilde{A}^{0,0}_{\bar{l}l}(l_{1},l_{2},l_{3},l_{4})
\rightarrow \tilde{A}^{k^{\prime},k}_{\bar{l}l}(l_{1},l_{2},l_{3},l_{4})]
\nonumber  \\
L^{k^{\prime},k}_{a_{e}a_{e}} & = & L^{0,0}_{v_{e}v_{e}}[\tilde{A}^{0,0}_{\bar{l}l}(l_{1},l_{2},l_{3},l_{4})
\rightarrow \tilde{A}^{k^{\prime},k}_{\bar{l}l}(l_{1},l_{2},l_{3},l_{4})]
\nonumber  \\
L^{k^{\prime},k}_{v_{e}a_{e}} & = & L^{0,0}_{v_{e}a_{e}}[\tilde{A}^{0,0}_{\bar{l}l}(l_{1},l_{2},l_{3},l_{4})
\rightarrow \tilde{A}^{k^{\prime},k}_{\bar{l}l}(l_{1},l_{2},l_{3},l_{4})]
\nonumber  \\
L^{k^{\prime},k}_{a_{e}v_{e}} & = & L^{0,0}_{a_{e}v_{e}}[\tilde{A}^{0,0}_{\bar{l}l}(l_{1},l_{2},l_{3},l_{4})
\rightarrow \tilde{A}^{k^{\prime},k}_{\bar{l}l}(l_{1},l_{2},l_{3},l_{4})]
\nonumber  \\
L^{0,k}_{v_{e}v_{e}} & = &  L^{0,0}_{v_{e}a_{e}}[\tilde{A}^{0,0}_{\bar{l}l}(l_{1},l_{2},l_{3},l_{4})
\rightarrow \tilde{A}^{0,k}_{\bar{l}l}(l_{1},l_{2},l_{3},l_{4})]
\nonumber  \\
L^{0,k}_{a_{e}a_{e}} & = &  L^{0,0}_{a_{e}v_{e}}[\tilde{A}^{0,0}_{\bar{l}l}(l_{1},l_{2},l_{3},l_{4})
\rightarrow \tilde{A}^{0,k}_{\bar{l}l}(l_{1},l_{2},l_{3},l_{4})]
\nonumber  \\
L^{0,k}_{v_{e}a_{e}} & = & L^{0,0}_{v_{e}v_{e}}[\tilde{A}^{0,0}_{\bar{l}l}(l_{1},l_{2},l_{3},l_{4})
\rightarrow \tilde{A}^{0,k}_{\bar{l}l}(l_{1},l_{2},l_{3},l_{4})]
\nonumber  \\
L^{0,k}_{a_{e}v_{e}} & = & L^{0,0}_{a_{e}a_{e}}[\tilde{A}^{0,0}_{\bar{l}l}(l_{1},l_{2},l_{3},l_{4})
\rightarrow \tilde{A}^{0,k}_{\bar{l}l}(l_{1},l_{2},l_{3},l_{4})].
\nonumber  \\
\end{eqnarray}
%\end{widetext}

Using the symmetry
\begin{equation}
\label{BB1}
A^{(I,I)}_{ll} (l_{1},l_{2},l_{3},l_{4}) =  A^{(I,I)}_{ll} (l_{3},l_{4},l_{1},l_{2}),
\end{equation}
and noting that the radial integrals are real, then
\begin{equation}
\label{BB2}
L^{0,0}_{v_{e}a_{e}} = L^{0,0}_{a_{e}v_{e}},
\end{equation}
and these terms cancel in (\ref{W68}).
Similarly, the symmetry
\begin{equation}
\label{BB3}
A^{(\lambda, \lambda)}_{\bar{l}l} (l_{1},l_{2},l_{3},l_{4}) =  
(-1)^{l+\bar{l}} A^{(-\lambda, -\lambda)}_{l\bar{l}} (l_{3},l_{4},l_{1},l_{2}),
\end{equation}
gives
\begin{equation}
\label{BB4}
L^{k,k}_{v_{e}a_{e}} = L^{k,k}_{a_{e}v_{e}} =0.
\end{equation}
Hence all the diagonal elements of $L^{(e)}_{fi}(\tilde{\mathbf{q}})^{\beta \alpha}$
are real.

\section{Evaluation of radial matrix elements}

The evaluation of the radial matrix elements (\ref{W53}) requires the integration
of the radial Dirac equations (\ref{W48}) to obtain the continuum eigenfunction 
\begin{equation}
\label{B1}
y_{\kappa_{f},E_{f}}(r) \equiv \left(\begin{array}{c} 
g_{\kappa_{f},E_{f}}(r) \\
f_{\kappa_{f},E_{f}}(r)
\end{array} \right),
\end{equation}
and the subsequent integration over a product of the initial bound state eigenfunction (\ref{W71}),
this continuum eigenfunction, and the spherical Bessel function $j_{l}(qr)$.  

The continuum eigenfunctions are regular at the origin and asymptotic to standing waves. 
These solutions increase as $r^{\gamma_{f}}$ until the point $r_{s} \approx |\kappa_{f}|/p$
where they become oscillatory with approximately constant amplitude. As $\gamma_{f}$ can 
be quite large, the smoothed functions~\cite{Pratt1964,Johnson1967}
\begin{equation}
\label{B2}
\mathcal{Y}_{\kappa_{f},E_{f}}(r) \equiv r^{-\gamma_{f}} \;y_{\kappa_{f},E_{f}}(r),
\end{equation}
are integrated outwards from $r=0$ until the first maximum or minimum of  
$\mathcal{Y}_{\kappa_{f},E_{f}}(r)$ is reached. These smoothed solutions satisfy
\begin{equation}
\label{B3}
\frac{d\mathcal{Y}_{\kappa_{f},E_{f}}(r)}{dr} = a(r,\gamma_{f}) \mathcal{Y}_{\kappa_{f},E_{f}}(r),
\end{equation}
where
\begin{equation}
\label{B4}
a(r, \gamma_{f}) = \left( \begin{array}{cc} 
- (\kappa_{f}+\gamma_{f})/r & E_{f}+m_{e}-V(r) \\
-(E_{f}-m_{e}-V(r)) & (\kappa_{f}-\gamma_{f})/r
\end{array} \right).
\end{equation}
The starting values were obtained from a series expansion about the origin
\begin{equation}
\label{B5}
\mathcal{Y}_{\kappa_{f},E_{f}}(r) = \sum_{n=0} \left( 
\begin{array}{c}
a_{n} \\
b_{n}
\end{array} \right) r^{n}  ,
\end{equation}
where the coefficients $a_{n}$ and $b_{n}$ ($n \ge 1$) satisfy the recurrence relations
\begin{eqnarray}
\label{B6}
a_{n} & = & [\alpha Z(m_{e}-E_{f})a_{n-1} \nonumber  \\
&&+ (m_{e}+E_{f})(n + \gamma_{f}-\kappa_{f})b_{n-1}]/n(2 \gamma_{f}+n),
\nonumber  \\
b_{n} & = & -[\alpha Z(m_{e}+E_{f})b_{n-1} \nonumber  \\
&&+ (E_{f}-m_{e})(n + \gamma_{f}+\kappa_{f})a_{n-1}]/n(2 \gamma_{f}+n),
\end{eqnarray}
with $a_{0}$ and $b_{0}$ determined from
\begin{equation}
\label{B7}
b_{0}=\left(\frac{\alpha Z}{\kappa_{f}-\gamma_{f}}\right) a_{0}, \quad (\kappa_{f} <0),
\end{equation}
and
\begin{equation}
\label{B8}
a_{0}=\left(\frac{\alpha Z}{\kappa_{f}+\gamma_{f}}\right) b_{0}, \quad \quad (\kappa_{f} >0).
\end{equation}
As the differential equations (\ref{B3}) are linear, the values $a_{0}=1$ ($\kappa_{f} < 0$)
and $b_{0}=1$ ($\kappa_{f}> 0$) may be used as initial conditions. However, the energy
normalization condition
\begin{equation}
\label{B9}
\int_{0}^{\infty} (g^{*}_{\kappa_{f}^{\prime}E_{f}^{\prime}} g_{\kappa_{f}E_{f}} +
f^{*}_{\kappa_{f}^{\prime}E_{f}^{\prime}} f_{\kappa_{f}E_{f}} ) dr = 
\delta_{\kappa_{f}^{\prime},\kappa_{f}} m_{e} \delta(E_{f}^{\prime}-E_{f})
\end{equation}
requires that the computed solutions be corrected by the factors~\cite{Johnson1967}
\begin{equation}
\label{B10}
a_{0} = 2 N_{f} \sqrt{E_{f}+m_{e}} (2p_{f})^{\gamma_{f}}\{ \gamma_{f} \cos \eta - y \sin \eta \},
\end{equation}
for $\kappa_{f}<0$, and
\begin{equation}
\label{B11}
b_{0} = -2 N_{f} \sqrt{E_{f}-m_{e}} (2p_{f})^{\gamma_{f}}\{ \gamma_{f} \sin \eta + y \cos \eta \},
\end{equation}
for $\kappa_{f}>0$.

The differential equations (\ref{B3}) were integrated using a fifth-order Runge-Kutta 
method~\cite{PTVF1992}. The rapid propagation of initial errors that arises from the 
$r^{-1}$ term in $a(r, \gamma_{f})$
was controlled by the use of the series expansion for the first $5|\kappa_{f}|$ points 
of the integration mesh~\cite{Johnson1967}. The integration algorithm correctly reproduced the 
free field solutions (\ref{W100}) for $Z=0$ and the asymptotic forms
\begin{equation}
\label{B12a}
\left ( \begin{array}{cc}
g_{\kappa_{f},E_{f}}(r) \\
f_{\kappa_{f},E_{f}}(r) 
\end{array}
\right)
=
\sqrt{\frac{m_{e}}{\pi p_{f}}} \left( \begin{array}{cc}
\sqrt{E_{f}+m_{e}} \;\cos (p_{f}r +\delta_{f}) \\
-\sqrt{E_{f}-m_{e}} \; \sin(p_{f}r+ \delta_{f})
\end{array}
\right), 
\end{equation}
for $Z \ne 0$, where the Coulombic phase is~\cite{Rose1961}
\begin{equation}
\label{B12b}
\delta_{f}=y_{f} \log (2p_{f}r)-\arg [\Gamma(\gamma_{f}+i y_{f})] + \eta_{f} - \pi \gamma_{f}/2.
\end{equation}

The radial integrals (\ref{W53}) have the form
\begin{equation}
\label{B12}
\mathcal{I}_{l}(q) = \int_{0}^{\infty} (\lambda_{i}r)^{\gamma_{i}+\beta_{i}} e^{-\lambda_{i}r} j_{l}(qr)
y_{\kappa_{f},E_{f}}(r) dr,
\end{equation}
where $\beta_{i}=0,1$, and were evaluated using an upper limit $r_{\infty}=x_{\infty}/\lambda_{i}$ 
with $x_{\infty}$ chosen to ensure $rj_{l}(qr)$ and $y_{\kappa_{f},E_{f}}(r)$ had attained their oscillatory forms. 
This required the condition
\begin{equation}
\label{B13}
r_{\infty} > \mathrm{max} \left(\frac{\sqrt{l(l+1)}}{q}, \frac{\sqrt{l_{f}(l_{f}+1)}}{p_{f}}, 
\frac{\sqrt{l_{f}^{\prime}(l_{f}^{\prime}+1)}}{p_{f}} \right)
\end{equation}
to be satisfied. Since $q_{\mathrm{min}}=\epsilon_{f}+\epsilon_{i}$, then,
for $\epsilon_{f}$ in the range~\cite{Gounaris2004} ($0.01 - 5.0$) keV, the smallest value of 
$q_{\mathrm{min}}$ occurs for H and is $4.6 \times 10^{-5}$. 
Since $p_{f} \approx \sqrt{2\epsilon_{f}}$  for small
$\epsilon_{f}$, then $p_{f,\mathrm{min}} \sim 6.3 \times10^{-3}$. Thus the condition on $j_{l}(qr)$
is the most challenging to meet.


\begin{thebibliography}{1}

\bibitem{Vergados2010} J.~D.~Vergados and Yu.~N.~Novikov, 
Exploring new features of neutrino oscillations with very low energy monoenergetic neutrinos, 
Nucl. Phys. B \textbf{839},1 (2010)

\bibitem{Thomas2016} A.~W.~Thomas and J.~D.~Vergados,
Solar neutrinos as background in dark matter searches involving electron detection,
J. Phys. G: Nucl. Part. Phys. \textbf{43}, 07LT01 (2016)

\bibitem{Campos2016} M.~D.~Campos and W.~Rodejohann,
Testing keV sterile neutrino dark matter in future direct detection experiments,
Phys. Rev. D \textbf{94}, 095010 (2016)

\bibitem{Giunti2015}
C.~Giunti and A.~Studenikin,
Neutrino electromagnetic interactions: A window to new physics,
Rev. Modern Phys. \textbf{87}, 531 (2015) 

\bibitem{Jeong2021}
J.~Jeong, J.~E.~Kim and S.~Youn,
Electromagnetic properties of neutrinos from scattering on bound electrons in atoms,
arXiv: 2105.01842 [hep-ph]

\bibitem{Kouzakov2014} K.~A.~Kouzakov and A.~I.~Studenikin,
Theory of neutrino-atom collisions: The history, present status, and BSM physics,
Advances in High Energy Physics \textbf{2014}, 569409 (2014)

\bibitem{Gaponov1976} Yu.~V.~Gaponov, Yu.~L.~Doprynin and V.~I.~Tikhonov, 
Elastic scattering of low energy neutrinos by atomic systems,
Sov. J. Nucl. Phys. \textbf{22}, 170 (1976)

\bibitem{Fayans1992} S.~A.~Fayans, V.~Yu.~Dobretsov and A.~B.~Dobrotsvetov, 
Effect of atomic binding on inelastic $\nu e$ scattering,
Phys. Lett. B \textbf{291}, 1 (1992)

\bibitem{Dobretsov1992} V.~Yu.~Dobretsov, A.~B.~Dobrotsvetov and S.~A.~Fayans, 
Inelastic neutrino scattering by atomic electrons,
Sov. J. Nucl. Phys. \textbf{55}, 1180 (1992)

\bibitem{Gounaris2002} G.~J.~Gounaris, E.~A.~Paschos and P.~I.~Porfyriadis, 
The ionization of H, He or Ne atoms using neutrinos or antineutrinos at keV energies,
Phys. Lett. B  \textbf{525}, 63 (2002)

\bibitem{Gounaris2004} G.~J.~Gounaris, E.~A.~Paschos and P.~I.~Porfyriadis,
Electron spectra in the ionization of atoms by neutrinos,
Phys. Rev. D \textbf{70}, 113008 (2004)

\bibitem{Voloshin2010} M.~B.~Voloshin,
Neutrino scattering on atomic electrons in searches for the neutrino magnetic moment,
Phys. Rev. Lett. \textbf{105}, 201801 (2010)

\bibitem{Kouzakov2011a} K.~A.~Kouzakov and A.~I.~Studenikin,
Magnetic neutrino scattering on atomic electrons revisited,
Phys. Lett. B \textbf{696}, 252 (2011)

\bibitem{Kouzakov2011b} K.~A.~Kouzakov, A.~I.~Studenikin, and M.~B.~Voloshin,
Neutrino-impact ionization of atoms in searches for neutrino magnetic moment,
Phys. Rev. D \textbf{83}, 113001 (2011)

\bibitem{Chen2013} 
J.-W.~Chen, C.-P.~Liu, C.-L.~Wu and C.-P.~Wu,
Ionization of hydrogen by neutrino magnetic moment, relativistic muon, and WIMP,
Phys. Rev. D \textbf{88}, 033006 (2013)

\bibitem{Chen2014}
J.-W.~Chen, H.-C.~Chi, K.-N.~Huang, C.-P.~Liu, H.-T.~Shiao, L.~Singh, H.~T.~Wong, C.-L.~Wu, and C.-P.~Wu,
Atomic ionization of germanium by neutrinos from \textit{ab initio} approach, Phys. Lett. B 
\textbf{731}, 159  (2014)

\bibitem{Chen2015}
Jiunn-Wei~Chen, Hsin-Chang~Chi, Keh-Ning~Huang, Hau~Bin~Li, C.-P.~Liu, Lakhwinder~Singh, Henry~T.~Wong, Chih-Liang~Wu, and Chih-Pan~Wu,
Constraining neutrino electromagnetic properties by germanium detectors,
Phys. Rev. D \textbf{91}, 013005  (2015)


\bibitem{Whittingham1971} I.~B.~Whittingham, Incoherent scattering of gamma rays in heavy atoms, 
J. Phys. A:Gen. Phys. \textbf{4}, 21 (1971)

\bibitem{Furry1951} W.~H.~Furry, On bound states and scattering in positron theory, Phys. Rev. \textbf{81}, 115 (1951)

\bibitem{IZ80} C.~Itzykson and J.~B.~Zuber,
\textit{Quantum Field Theory} (New York, McGraw-Hill, 1980)

\bibitem{Bailin82} D.~Bailin, 
\textit{Weak Interactions} 2nd ed. (Bristol, Adam Hilger, 1982)

\bibitem{Jauch55} J.~M.~Jauch and F.~Rohrlich,
\textit{The Theory of Photons and Electrons} (Reading, Addison-Wesley, 1955)

\bibitem{Rose1961} M.~E.~Rose \textit{Relativistic Electron Theory} (New York, Wiley, 1961)

\bibitem{Olsen1955} H.~Olsen,
Outgoing and ingoing waves in final states and bremsstrahlung,
Phys. Rev. \textbf{99}, 1335 (1955)

\bibitem{Rose1957} M.~E.~Rose \textit{Elementary Theory of Angular Momentum} (New York, Wiley, 1957)

\bibitem{Bethe1957} H.~A.~Bethe and E.~E.~Salpeter \textit{Quantum Mechanics of One- 
and Two- Electron Atoms} (Berlin, Springer-Verlag, 1957)

\bibitem{Thomas1997} D.~Thomas \texttt{http://www.chembio.uoguelph.ca \\
/educmat.atomdata/shield/grp18nsh.htm}

\bibitem{Schofield1973} J.~H.~Schofield,
Theoretical photoionization cross sections from 1 to 1500 keV,
Lawrence Livermore Laboratory report UCRL-51326 (1973),
\texttt{https://doi.org/10.2172/4545040}.

\bibitem{Wimp1981} J.~Wimp,
\textit{Sequence Transformations and Their Applications},
(New York, Academic Press, 1981), p.169

\bibitem{Pratt1964}
R.~H.~Pratt, R.~D.~Levee, R.~L.~Paxton, and W.~Aron, $K$-Shell Photoelectric Cross Sections
from 200 keV to 2 MeV, Phys. Rev. \textbf{134}, A898 (1964)

\bibitem{Johnson1967} W.~R.~Johnson,
Angular distribution of single-quantum annihilation radiation,
Phys. Rev. \textbf{159}, 61 (1967)

\bibitem{PTVF1992} W.~H.~Press, S.~A.~Teukolsky, W.~T.~Vetterling and B.~P.~Flannery,
\textit{Numerical Recipes in Fortran}, Second Ed. (Cambridge, Cambridge University Press, 1992)



\end{thebibliography}
\end{document}